\documentclass[pra,10pt,showpacs,notitlepage]{revtex4-1}
\usepackage{amsmath}    
\usepackage{graphicx}   
\usepackage{verbatim}   
\usepackage{color}      
\usepackage{subfigure}  
\usepackage{hyperref}   

\begin{document}

\title{Sound waves and modulational instabilities on continuous wave solutions in spinor
Bose-Einstein condensates}

\author{Richard S. Tasgal and Y. B. Band}

\affiliation{Department of Chemistry, Department of Physics,
Department of Electro-Optics, and the Ilse Katz Center for Nano-Science,
Ben-Gurion University, Beer-Sheva 84105, Israel}

\begin{abstract}
We analyze sound waves (phonons, Bogoliubov excitations) propagating
on continuous wave (cw) solutions of repulsive $F=1$ spinor Bose-Einstein
condensates (BECs), such as $^{23}$Na (which is anti-ferromagnetic or
polar) and $^{87}$Rb (which is ferromagnetic).
Zeeman splitting by a uniform magnetic field is included.  
%
All cw solutions to ferromagnetic BECs with vanishing $M_F=0$ particle
density and non-zero components in both $M_F=\pm 1$ fields are subject
to modulational instability (MI).
Modulational instability increases with increasing particle density.
Modulational instability also increases with differences in the components' wave numbers;
this effect is larger at lower densities
but becomes insignificant at higher particle densities.
Continuous wave solutions to anti-ferromagnetic (polar) BECs with vanishing $M_F=0$
particle density and non-zero components in both $M_F=\pm 1$ fields do
not suffer MI if the wave numbers of the components are the same.  If
there is a wave number difference, MI initially increases with
increasing particle density, then peaks before dropping to zero beyond
a given particle density.
%
The cw solutions with particles in both $M_F=\pm 1$ components and non-vanishing
$M_F=0$ components do not have MI if the wave numbers of the
components are the same, but do exhibit MI when the wave numbers are
different.
Direct numerical simulations of a cw with weak white noise confirm
that weak noise grows fastest at wave numbers with the largest MI,
and show some of the results beyond small amplitude perturbations.
Phonon dispersion curves are computed numerically; we find analytic
solutions for the phonon dispersion in a variety of limiting cases.
\end{abstract}


\maketitle

\section{Introduction}  \label{Sec:introduction}

Bose-Einstein condensates (BECs) \cite{Bose.1924, Einstein.1924, Einstein.1925, Anderson.1995, Davis.1995} hold the 
promise of opening many new vistas in physics, e.g., macroscopic systems that exhibit quantum effects,
higher resolution measurements of time, inertia, and other quantities, and a medium with which to carry out quantum 
computing and to simulate quantum systems 
\cite{Feynman.1986, CiracZoller.2012, Lahav.2010, Gorshkov.2010, Lewenstein.2012}.
Many interesting phenomena in BECs either occur \textit{against} simpler backgrounds or are prepared 
\textit{from} initially simpler states, which are often plane waves or approximations thereof.
For example, vortices \cite{Saito.2007, Lamacraft.2007, UhlmannSchutzholdFischer.2007, KawaguchiUeda.2012} 
and dark solitons \cite{Burger.1999, Denschlag.2000, Li.2005} are typically imbedded on plane waves,
and a spin texture \cite{Sadler.2006} may be composed of, in part, many regions that are approximately plane 
waves.  When the dynamics of plane waves are understood better, the structures that sit on them may be 
understood better.  It is useful to know---especially if those simpler states are not quite as simple as had 
been thought---when they can and cannot exhibit more complex dynamics, and what those dynamics are.

Bose-Einstein condensates can be composed of particles with
non-zero total angular momentum ($F>0$). For example, there is $H$ \cite{Fried.1998},
$^7$Li \cite{Bradley.1997}, 
$^{23}$Na \cite{Davis.1995, Simkin.1999}, 
$^{41}$K \cite{Modugno.2001}, 
$^{52}$Cr \cite{Griesmaier.2005}, 
$^{84}$Sr \cite{Stellmer.2009},
$^{85}$Rb \cite{Cornish.2000},  
$^{87}$Rb \cite{Anderson.1995}, 
$^{133}$Cs \cite{Weber.2003},
$^{164}$Dy \cite{Lu.2011}, and $^{170}$Yb \cite{Fukuhara.2007}.
An optical (as opposed to magnetic) trap can hold all the spin
components ($M_F=-F, -F+1, \ldots, F-1, F$) \cite{Barrett.2001, Chang.2004, Beaufils.2008, Leslie.2009}.  
In this case the BEC field is a spinor with $2F + 1$ $M_F$ components, and can exhibit
phenomena that do not occur in scalar fields.
First and foremost, there is magnetism.  More complicated phenomena in
spinor BECs that do not occur in a scalar BEC are some forms of MI 
\cite{Robins.2001, Konotop.2002, Li.2005, Zhang.MI.2005, Kunimi.2014},
oscillatory coherent spin mixing \cite{Zhang.spin_mixing.2005, MurPetit.2009,TasgalBand.2013},
formation of spin textures, i.e., patterns of spatial variation of the magnetization \cite{Sadler.2006},
and certain forms of vortices with magnetization \cite{Sadler.2006, Saito.2007, Lamacraft.2007,UhlmannSchutzholdFischer.2007},
including fractional vortices and non-Abelian vortices \cite{KawaguchiUeda.2012}.

Here we examine sound waves (phonons, acoustic waves, Bogoliubov
excitations) that propagate on top of continuous wave (cw) solutions
of $F=1$ spinor BECs \cite{Mewes.1996, Andrews.1997, Andrews.1998,
Steinhauer.2002, Ozeri.2005, Bogolubov.1947}.
When the frequencies of the sound waves have (do not have) imaginary
parts, they grow exponentially (do not grow), which implies that the background is unstable (stable).
Sound waves with imaginary parts are most often called MI or Benjamin-Feir
instabilities  \cite{BespalovTalanov.1966, BenjaminFeir.1967, Agrawal.2001};
there are instances where it has been called self-pulsing instability \cite{Agrawal.2001} or
dynamical instability \cite{KawaguchiUeda.2012}.
%
%
Section~\ref{Sec:model} introduces
the equations for the dynamics of an $F=1$ spinor BEC, the general form
of the cw solutions, and sets out the formalism for describing small
amplitude sound waves.  Section~\ref{Sec:band_diagrams} computes the
phonon band diagrams, i.e., the way the frequencies (chemical
potentials) of the sound waves depend on wave number.  Special
attention is paid to complex-valued frequencies, since this, MI,
causes the sound waves to grow exponentially.
%
%
Section~\ref{Sec:MI_evolution} presents direct numerical simulations,
which confirm the analytic results for small amplitude phonons, and
investigates the evolution of large amplitude (highly nonlinear)
noise.  
%
%
%
Section~\ref{Sec:conclusions} summarizes.

\section{Quantitative model for spinor BECs with magnetic fields}
\label{Sec:model}

The Hamiltonian density for an $F=1$ spinor BEC with linear and
quadratic Zeeman splittings induced by a magnetic field ${\bf B} = 
B \, \hat {\bf z}$ (and without spin-dipolar coupling) is
\cite{OhmiMachida.1998, Ho.1998, StamperKurnUeda.2013}
\begin{eqnarray}
\mathcal{H}
& = & \frac{\hbar^2}{2 m} \mathbf{\nabla} \Phi_a^\dagger \cdot \mathbf{\nabla}  \Phi_a 
   +     \frac{c_0}{2} \Phi_a^\dagger \Phi_b^\dagger \Phi_b \Phi_a
   +     \frac{c_2}{2} \Phi_a^\dagger \Phi_{a'}^\dagger \mathbf{F}_{ab} \cdot \mathbf{F}_{a'b'} \Phi_{b'} \Phi_b
   -      p B \Phi_a^\dagger F^z_{ab} \Phi_b
   +     q B^2  \Phi_a^\dagger (F^z_{ab})^2 \Phi_b \, .
\end{eqnarray}
Here $\Phi = (\phi_1,\phi_0,\phi_{-1})^\mathrm{t}$ is a
vector composed of the amplitudes of spin $M_F=1$, $M_F=0$, and $M_F=-1$.
$m$ is the mass of the particles, $c_0$ and $c_2$
are the coefficients of the spin-independent and spin-dependent parts
of the mean field, ${\bf F}$ is the total atomic angular momentum
vector and each component is a 3$\times$3
spin-1 matrix; $B$ is the magnitude of
the (uniform) external magnetic field, which is taken to be in the
$z$-direction, and $p$ and $q$ are linear and quadratic Zeeman
coefficients \cite{Saito.2007, Lamacraft.2007, Zhang.spin_mixing.2005}.
The results here actually apply to any strength Zeeman effect,
and are not limited to magnetic fields small enough
for the dependence of the particle energies as a function of $B$
to be described by the first two terms of a Taylor expansion.
To generalize, substitute relative particle energies as a function of spin state
$[E_{m=+1}(B) - E_{m=-1}(B)]/2$ for $p B$ and 
$[E_{m=+1}(B) + E_{m=-1}(B) - 2 E_{m=0}(B)]/2$ for $q B^2$.
Within the limitations of this paper, $[E_{m=+1}(B) + E_{m=0}(B) + E_{m=-1}(B)]/3$
is a global energy shift, and may be neglected.

If the BEC is in a quasi-one-dimensional optical trap with population only in a
single transverse bound state, the governing equations are
\begin{subequations}
\label{Eqs:SpinorBEC}
\begin{eqnarray}
i \hbar \frac{\partial}{\partial t} \phi_1 
& = & -\frac{\hbar^2}{2 m} \frac{\partial^2}{\partial z^2} \phi_1
           + c_0 \left( |\phi_1|^2 + |\phi_0|^2 + |\phi_{-1}|^2 \right) \phi_1
           + c_2 \left[ \left( |\phi_1|^2 + |\phi_0|^2 - |\phi_{-1}|^2 \right) \phi_1 
	                  + \phi_0^2 \phi_{-1}^*
	             \right]  \nonumber \\
& & \;\;\;\;\;\;\;\;\;\;\;\;\;\;\;\;\;\;\;\; + (-p B + q B^2) \phi_1 \, ,
      \label{Eq:phi+1} \\
i \hbar \frac{\partial}{\partial t} \phi_0 & = &
           - \frac{\hbar^2}{2 m} \frac{\partial^2}{\partial z^2} \phi_0
           + c_0 \left( |\phi_1|^2 + |\phi_0|^2 + |\phi_{-1}|^2 \right) \phi_0        
           + c_2 \left[ \left( |\phi_1|^2 + |\phi_{-1}|^2 \right) \phi_0
                            + 2 \phi_1 \phi_0^* \phi_{-1}
                      \right] \, ,
      \label{Eq:phi0} \\
i \hbar \frac{\partial}{\partial t} \phi_{-1} 
& = & -\frac{\hbar^2}{2 m} \frac{\partial^2}{\partial z^2} \phi_{-1}
           + c_0 \left( |\phi_1|^2 + |\phi_0|^2 + |\phi_{-1}|^2 \right) \phi_{-1} 
+ c_2 \left[ \left( -|\phi_1|^2 + |\phi_0|^2 + |\phi_{-1}|^2 \right) \phi_{-1}
                            + \phi_0^2 \phi_1^*
                      \right] \nonumber \\
& & \;\;\;\;\;\;\;\;\;\;\;\;\;\;\;\;\;\;\;\;\;\; + (p B + q B^2) \phi_{-1} \, .
      \label{Eq:phi-1}
\end{eqnarray}
\end{subequations}
Time and space are $t$ and $z$, $c_0$ is a coefficient of self-phase modulation,
and $c_2$ is a coefficient of both self-phase modulation
and parametric nonlinearity (and is spin-dependent).
Materials with negative $c_2$ are ferromagnetic,
and materials with positive $c_2$ are anti-ferromagnetic, or polar. 
%
The nonlinear coefficients are functions of the particle mass $m$
and the s-wave scattering lengths $a_0$, $a_2$ for the $F=0$ and $F=2$
channels, $g_0 = (4\pi\hbar^2/m) a_0$, $g_2 = (4\pi\hbar^2/m) a_2$,
with the nonlinear coefficients in the governing equations above $c_0
= (g_0 + 2 g_2) / 3$, $c_2 = -(g_0 - g_2) / 3$.
The values of the nonlinear coefficients are modified when the BEC is
in a trap \cite{Olshanii.1998, Bergeman.2003}.
%
%
The scattering lengths of $^{87}$Rb are $a_0 = 101.8\,a_B$ and $a_2 =
a_0 - 1.45 \, a_B$, where $a_B$ is the Bohr radius \cite{Chang.2005,
KlausenBohnGreene.2001, KempenKokkelmansHeinzenVerhaar.2002},
%
%
and the scattering lengths of $^{23}$Na have been measured to be $a_0
= 50.0\,a_B$, $a_2 = a_0 + 5.0\,a_B$.
%
%
%
%
The ratios $c_2/c_0$ are $-0.0048$ for $^{87}$Rb and $0.031$ for $^{23}$Na.
%
%
%
%
Equations~(\ref{Eqs:SpinorBEC}) are integrable when $c_2=0$ (in which
case the system is a set of generalized Manakov equations
\cite{Nakkeeran.1998, Manakov.1973}) or $c_2=c_0$
\cite{IedaMiyakawaWadati.2004, WadatiTsuchida.2006, IedaWadati.2007}.

Equations~(\ref{Eqs:SpinorBEC}) can be written in dimensionless form
by applying a change of variables
\begin{subequations}
\label{nondimensionalization}
\begin{eqnarray}
t' & = & t / t_d , \\
z' & = & z / z_d = z / \sqrt{\hbar t_d / m} , \\
\phi_j' & = & \phi_j / \phi_d = \phi_j / \sqrt{\hbar / (c_0 t_d)} .
\end{eqnarray}
\end{subequations}
The dimensionless equations have $\hbar=1$, $m=1$, $c_0=1$, and
$c_2|_\mathrm{dimensionless}=c_2/c_0$.  The dimensionless frequencies
and wave numbers go as the dimensional variables times $t_d$ and
$z_d=\sqrt{\hbar t_d / m}$, respectively.  $t_d$ is a free variable,
and may be chosen such that the dimensionless time, space, or
amplitudes are convenient magnitudes.  We will use dimensionless
variables in the figures in order to emphasize generality, but retain
the dimensions in the body of the text to more closely connect the
equations to the physical parameters.

A important observable is the magnetization vector $\mathbf{m}=(m_x, m_y, m_z)$.
For a spin $F = 1$ BEC, this is the spin-vector density,
which is equal to the expectation value of the spin-vector $\mathbf{F} = (F_x, F_y, F_z)$, where
\begin{equation}
F_x = \frac{1}{\sqrt{2}} \left( \begin{matrix} 0 & 1 & 0 \\ 1 & 0 & 1 \\ 0 & 1 & 0 \end{matrix} \right) , \;
F_y = \frac{ 1}{\sqrt{2}} \left( \begin{matrix} 0 & -i  & 0 \\ i & 0 & -i \\ 0 & i & 0 \end{matrix} \right) , \;
F_z = \left( \begin{matrix} 1 & 0 & 0 \\ 0 & 0 & 0 \\ 0 & 0 & -1 \end{matrix} \right) , 
\end{equation}
multiplied by the maximum magnetic moment of the particles that constitute the BEC.
For the dimensionless variables, we take the maximum magnetic moment to be unity.
That gives the dimensionless magnetization vector
\begin{equation}
\mathbf{m} 
= \left( \begin{matrix} m_x \\ m_y \\ m_z \end{matrix} \right) 
= \left( \begin{matrix}
     \mathrm{Re}[ \sqrt{2} (\phi_1 + \phi_{-1})^* \phi_0 ] \\ 
     \mathrm{Im}[ \sqrt{2} (\phi_1 - \phi_{-1})^* \phi_0 ] \\ 
     |\phi_1|^2 - |\phi_{-1}|^2
   \end{matrix} \right) .
\end{equation}

If any two of the three spin fields are zero, then the remaining field is
governed by a simple nonlinear Schr\"{o}dinger (NLS) equation, called the 
Gross--Pitaevskii equation, which is completely integrable 
\cite{ZakharovShabat.1971, ZakharovShabat.1973}.
If the $M_F=0$ component spin field vanishes ($\phi_0=0$), then the spin
$M_F=\pm 1$ fields ($\phi_1$, $\phi_{-1}$) are governed by a pair of
coupled nonlinear Schr\"{o}dinger (CNLS) equations,
\begin{subequations}
\label{CNLS:SpinorBEC}
\begin{eqnarray}
i \hbar \frac{\partial}{\partial t} \phi_1 & = &
         - \frac{\hbar^2}{2 m} \frac{\partial^2}{\partial z^2} \phi_1
         + \left[ (c_0 + c_2) |\phi_1|^2 + (c_0 - c_2) |\phi_{-1}|^2 \right] 
	 \phi_1 \, ,
      \label{CNLS:phi+1} \\
i \hbar \frac{\partial}{\partial t} \phi_{-1} & = &
         - \frac{\hbar^2}{2 m} \frac{\partial^2}{\partial z^2} \phi_{-1}
         + \left[ (c_0 - c_2) |\phi_1|^2 + (c_0 + c_2) |\phi_{-1}|^2 \right] 
	 \phi_{-1} \, ,
      \label{CNLS:phi-1}
\end{eqnarray}
\end{subequations}
which have been intensely studied (see, e.g., Ref.~\cite{Yang.2010}).
They describe, among other physical systems, light in optical fibers \cite{Agrawal.2001}.
In optics, one convention is to describe the ratio of the coefficients of cross- to self-phase modulation
as $B=(c_0-c_2)/(c_0+c_2)$. Linear polarization in optics is described by $B=2/3$,
which corresponds to a ratio $c_2/c_0=1/5$; circular polarization in optics has
$B=2$, which corresponds to $c_2/c_0=-1/3$. Because of the positive and negative
signs of the corresponding  $c_2/c_0$, the dynamics of linearly polarized light
resembles the dynamics of repulsive ferromagnetic BECs, and circularly
polarized light resembles anti-ferromagnetic (polar) BECs, provided the BECs do not
contain any particles with spin $m=0$. The CNLS equations for light in a fiber often contain
group-velocity birefringence terms. These terms may be eliminated by a transformation that 
changes variables: shifts the frequencies and wave numbers of the component fields [see, e.g., 
Eq.~(7.2.29) in~\cite{Agrawal.2001}]. Changing the group-velocity birefringence (e.g.,  setting it to zero)
corresponds to shifting the wave numbers of the spin $m=+1,-1$ up and down, respectively.
%
%

\subsection{Continuous wave solutions}
\label{Subsec:cw_solutions}

We begin by considering the solutions of the dynamical equations with
the simplest shape, a flat constant field, or cw. 
The \textit{most general} cw ansatz is
\begin{subequations}
\label{cw.ansatz}
\begin{eqnarray}
\phi_1 & = & A_1 \exp[i (\theta_1 + k_1 z - \omega_1 t)]  \, ,
      \label{cw.ansatz.phi+1} \\
\phi_0 & = & A_0 \exp[i (\theta_0 + k_0 z - \omega_0 t)] \, ,
       \label{cw.ansatz.phi0} \\
\phi_{-1} & = & A_{-1} \exp[i (\theta_{-1} + k_{-1} z - \omega_{-1} t)] \, ,
      \label{cw.ansatz.phi-1}
\end{eqnarray}
\end{subequations}
where the parameters are real-valued and, without loss of generality,
$A_j$ are positive definite.
%
%
Spin components with different wave numbers have, in general, different
velocities. In this model, the domain is infinite.  In an
experiment (or in numerical simulations), a BEC with different
wave numbers in the different spin components would need to be long
enough to avoid edge effects, or one could replenish the fields at the
boundaries, or arrange the fields in a ring (confined by a toroidal
potential) \cite{Gupta.2005, Olson.2007, Lesanovsky.2007,
Henderson.2009, Ryu.2007}. In a ring, the wave numbers would be
quantized, but otherwise all the results would hold. Modulation
instability of spinor BECs in a ring geometry has been studied
theoretically in Ref.~\cite{Kunimi.2014}.

Let us substitute the cw trial function~(\ref{cw.ansatz}) into the dynamical
equations~(\ref{Eqs:SpinorBEC}).
If $c_2 = 0$, there are cw solutions for every value of the amplitude
($A_j$), wave number ($k_j$), and phase ($\theta_j$).  The frequencies
of the fields are
\begin{equation}
\hbar \omega_j = 
\frac{\hbar^2 k_j^2}{2 m} + c_0 (A_1^2 + A_0^2 + A_{-1}^2) \, .
\end{equation} 
If $c_2 \neq 0$, then the parametric term requires a relation between
the phases of the three fields,
\begin{subequations}
\label{cw.ansatz.phases.0}
\begin{eqnarray}
k_0 & = & \frac{1}{2} (k_1 + k_{-1}) \, , \label{cw.ansatz.k.0} \\
\omega_0 & = & \frac{1}{2} (\omega_1 + \omega_{-1}) \, , 
\label{cw.ansatz.omega.0} \\
\theta_0 & = & \frac{1}{2} (\theta_1 + \theta_{-1} + n \pi) \, , 
\label{cw.ansatz.theta.0} 
\end{eqnarray}
\end{subequations}
where $n$ is an integer.
The frequencies of the spin components $M_F=1,0,-1$ are
\begin{subequations}
\label{cw.magnitudes}
\begin{eqnarray}
\hbar \omega_1 & = & \frac{\hbar^2 k_1^2}{2 m}
           + c_0 \left( A_1^2 + A_0^2 + A_{-1}^2 \right)
           + c_2 \left[ A_1^2 + A_0^2 - A_{-1}^2 + (-1)^n A_0^2 A_{-1} / 
	   A_1 \right]
	  + (-p B + q B^2) A_1 \, ,
      \label{cw.A+1} \\
\hbar \omega_0 & = & \frac{\hbar^2 k_0^2}{2 m}
           + c_0 \left( A_1^2 + A_0^2 + A_{-1}^2 \right)
           + c_2 \left[ A_1^2 + A_{-1}^2 + 2 (-1)^n A_1 A_{-1} \right] \, ,
      \label{cw.A_0} \\
\hbar \omega_{-1} & = & \frac{\hbar^2 k_{-1}^2}{2 m}
           + c_0 \left( A_1^2 + A_0^2 + A_{-1}^2 \right)
           + c_2 \left[ -A_1^2 + A_0^2 + A_{-1}^2 + (-1)^n A_0^2 A_1 / 
	   A_{-1} \right] 
	  + (p B + q B^2) A_{-1} \, .
      \label{cw.A-1}
\end{eqnarray}
\end{subequations}
For consistency of the frequency $\omega_0$ of the $M_F=0$ field,
[Eqs.~(\ref{cw.magnitudes}) and~(\ref{cw.ansatz.omega.0})],
the magnitude $A_0$ of the $M_F=0$ field must be
\begin{equation}
A_0^2 = 2 (-1)^n A_1 A_{-1}
                \left(
                  1 - \frac{ (\hbar^2 / 2 m) [(k_1 - k_{-1})/2]^2 + q B^2}
                               {c_2 [A_1 + (-1)^n A_{-1}]^2}
                \right) .
\label{cw.A_0.solution}
\end{equation}
The left hand side of Eq.~(\ref{cw.A_0.solution}) is real and
non-negative.  If $c_2<0$ (ferromagnetic), for the right hand side to
be positive, the cw solutions must have even $n$ (for conciseness, we
write $n=0$ to denote even $n$ and $n=1$ for odd $n$).  If $c_2>0$
(anti-ferromagnetic), then over certain ranges of the cw
spin-components $M_F=\pm 1$, there are only $n=0$ solutions, both
$n=0$ and $n=1$ solutions, or only $n=1$ solutions; there is always at
least one cw solutions with non-vanishing $M_F=0$ field, with the
exception of $M_F=\pm 1$ particle densities exactly equal to each
other and below the threshold at which $n=0$ cws exist, $A_1^2 =
A_{-1}^2 < (4 c_2)^{-1} [\hbar^2 (k_1 - k_{-1})^2 / 8 m + q B^2]$.
For the $n=0$ and $n=1$ type cw solutions, there is a particle
density $A_\mathrm{ref}$ which roughly separates
the regimes in which linear components of the energy are more
important from regimes in which nonlinear polarization-dependent
components of the energy are more important.
Taking the internal (in the reference frame in which the spin $m=0$ field has wave number zero) kinetic energy 
plus the quadratic Zeeman energy, i.e., all the linear terms, to be
$E_\mathrm{lin} = \{ \hbar^2 [(k_1 - k_{-1})/2]^2 / (2 m) + q B^2\} |A_\mathrm{ref}|^2$ 
and the energy scale of the nonlinear polarization-dependent terms as 
$E_\mathrm{NL} = c_2 |A_\mathrm{ref}|^4$, 
the particle density at which the absolute values of these linear and nonlinear energies are equal is 
$A_\mathrm{ref}^2 = | \{ \hbar^2 [(k_1 - k_{-1})/2]^2 / (2 m) + q B^2 \} / c_2|$. 
This reference particle density is a rough marker for the blurred boundary between the two regimes.
See Ref.~\cite{TasgalBand.2013} for a more detailed elucidation
of the consequences of Eq.~(\ref{cw.A_0.solution}).
The dimensionless magnetization components of this cw are
\begin{subequations}
\begin{eqnarray}
m_x & = & \mathrm{Re}\{ (i)^n \sqrt{2} A_0 [ A_1 \exp( i \phi_\Delta / 2 ) + A_{-1} \exp( -i \phi_\Delta / 2 ) ] \} , \\
m_y & = & \mathrm{Im}\{ (i)^n \sqrt{2} A_0 [ A_1 \exp( i \phi_\Delta / 2 ) - A_{-1} \exp( -i \phi_\Delta / 2 ) ] \} , \\
m_z & = & |A_1|^2 - |A_{-1}|^2 ,
\end{eqnarray}
\end{subequations}
where the phase difference between the $+1$ and $-1$ components is
$\phi_\Delta \equiv \phi_1 - \phi_{-1} + (k_1-k_{-1}) z - (\omega_1 - \omega_{-1}) t$.
The magnetization is flat in the ``natural'' orientation direction (z) of the cw and sinusoidal in space and time
in the transverse directions.

\subsection{Sound waves and modulational instabilities}
\label{Subsec:sound}

Sound waves in BECs have been studied experimentally in, e.g.,
Refs.~\cite{Mewes.1996, Andrews.1997, Andrews.1998, Steinhauer.2002, Ozeri.2005}. 
In the context of mean field theory, sound waves (or by other names, acoustic waves, phonons, or
Bogoliubov excitations) and also MI (where sound waves have
complex-valued frequencies) may be represented by small perturbations
to a cw solution \cite{Bogolubov.1947, BespalovTalanov.1966, BenjaminFeir.1967,
Menyuk.1987, Agrawal.1987, Agrawal.2001, Berman.2002, Robins.2001, HuangLiSzeftel.2004, Li.2005, Zhang.MI.2005, Kunimi.2014},
\begin{equation}
\phi_j = [A_j + a_j(z,t)] \exp[i (\theta_j + k_j z - \omega_j t)]  \, ,
\label{MI.ansatz}
\end{equation}
where $j=1,0,-1$.  It is convenient to define the frequencies and wave
numbers of the phonons with respect to the cw solution on which it
propagates, rather than $\phi_j = A_j \exp[i (\theta_j + k_j z -
\omega_j t)] + a_j(z,t)$.

\section{Phonon dispersion band diagrams}
\label{Sec:band_diagrams}

The dynamics of sound waves on top of the cw solutions are obtained by
substituting the ansatz~(\ref{MI.ansatz}) into the governing equations
(\ref{Eqs:SpinorBEC}).  If the sound waves are weak, one may linearize
in the perturbations $a_j$.
The perturbations then have superposition, and the general solution is
a sum of sound waves.  This allows a spectral approach.  Look for
solutions one frequency and wave number at a time, i.e., eigenvalues
and eigenvectors,
\begin{equation}
a_j(z,t) = p_j \cos(k z - \omega t) + i q_j \sin(k z - \omega t) .
\end{equation}
The phonons are here represented in terms of sines and cosines, but
they could equally well be in terms of exponentials.  The former tends
to be use more often in looking for MI (frequencies with complex
values) \cite{Agrawal.2001}, while the latter is more typical when
considering stable sound waves (Bogoliubov excitations) or more
quantum mechanical problems.

There are six equations in $p_j$ and $q_j$, each of which are
complex-valued,
\begin{eqnarray}
0 = M v = \left\{ [-\hbar \omega 
                       + \frac{\hbar^2}{2 m} (k_1 + k_{-1}) k] \mathbf{I} 
                       + \mathbf{P} 
                       + \mathbf{Q} 
                       + \mathbf{R} 
                 \right\} v \, ,
\label{phonon:eigenproblem}
\end{eqnarray}
where $v = (p_1, q_1, p_0, q_0, p_{-1}, q_{-1})^t$ and
\begin{subequations}
\label{phonon:eigenproblem:vectors}
\begin{equation}
\mathbf{P} =
\frac{\hbar^2 k}{2 m} 
\left(
  \begin{array}{cccccc}
    \Delta k & k & 0 & 0 & 0 & 0 \\
    k & \Delta k & 0 & 0 & 0 & 0 \\ 
    0 & 0 & 0 & k & 0 & 0 \\
    0 & 0 & k & 0 & 0 & 0 \\ 
    0 & 0 & 0 & 0 & -\Delta k & k\\
    0 & 0 & 0 & 0 & k & -\Delta k
\end{array}
\right)
\end{equation}
\begin{equation}
\mathbf{Q} =
2
\left(
  \begin{array}{cccccc}
    0 & 0 & 0 & 0 & 0 & 0 \\
    (c_0 + c_2) A_1^2 & 0 & (c_0 + c_2) A_1 A_0 & 0 & (c_0 - c_2) 
    A_1 A_{-1} & 0 \\ 
    0 & 0 & 0 & 0 & 0 & 0 \\
    (c_0 + c_2) A_1 A_0 & 0 & c_0 A_0^2 & 0 & (c_0 + c_2) A_0 A_{-1} & 0 \\ 
    0 & 0 & 0 & 0 & 0 & 0 \\
    (c_0 - c_2) A_1 A_{-1} & 0 & (c_0 + c_2) A_0 A_{-1} & 0 & (c_0 + c_2) 
    A_{-1}^2 & 0
  \end{array}
\right) ,
\end{equation}
\begin{equation}
\mathbf{R} =
(-1)^n c_2
\left(
  \begin{array}{cccccc}
    0 & -A_0^2 A_{-1} / A_1 & 0 & 2 A_0 A_{-1} & 0 & -A_0^2 \\
    -A_0^2 A_{-1} / A_1 & 0 & 2 A_0 A_{-1} & 0 & A_0^2 & 0 \\ 
    0 & 2 A_0 A_{-1} & 0 & -4 A_1 A_{-1} & 0 & 2 A_1 A_0 \\
    2 A_0 A_{-1} & 0 & 0 & 0 & 2 A_1 A_0 & 0 \\ 
    0 & -A_0^2 & 0 & 2 A_1 A_0 & 0 & -A_0^2 A_1 / A_{-1} \\
    A_0^2 & 0 & 2 A_1 A_0 & 0 & -A_0^2 A_1 / A_{-1} & 0
  \end{array}
\right) .
\end{equation}
\end{subequations}
Note that the frequency (chemical potential) $\omega$ of the
perturbations (relative to the cw on which it sits) and the average
wave number $k_0 = (k_1+k_{-1})/2$ appear only on the diagonal.
Off-diagonal terms depend on the parameters of the cw, $A_1$,
$A_{-1}$, $\Delta k \equiv k_1 - k_{-1}$, and $n$; note that $A_0$ is
a function of the other cw parameters [Eq.~(\ref{cw.A_0.solution})].
$c_2/c_0$ cannot be simplified or reduced to more trivial terms. The
linear and quadratic Zeeman effects do not appear explicitly.  The
quadratic Zeeman terms appear implicitly in the magnitude of $A_0$.
The quadratic Zeeman splitting will affect the perturbations of the cw
solutions with non-zero $M_F=0$ fields, but will not affect the cases
without an $M_F=0$ field.

Equations~(\ref{phonon:eigenproblem})--(\ref{phonon:eigenproblem:vectors})
constitute an eigenvalue--eigenvector problem, with solutions being
six eigenvalues $\omega_j=\omega_j(k)$, with eigenvectors
$(p_{j,i}(k),q_{j,i}(k))$, $j=1, \ldots, 6$. 
Some insights into the phonon dispersion curves may be obtained by
expanding the characteristic polynomial $|M|$ from
Eqs.~(\ref{phonon:eigenproblem})--(\ref{phonon:eigenproblem:vectors}),
in energy $\hbar\omega$ and wave number $k$.  We may express the
sixth-order polynomial equation for the eigenvalues (frequencies or
energies of the sound waves or Bogoliubov excitations), leaving the
dependence on the background cw implicit in the coefficients,
\begin{eqnarray} 
\label{Eq.phonon_dispersion_curve}
0 & = & (\hbar\omega)^6
         + ( C_{k^0\omega^4}^{\mathrm{cw}(n=0,1)}
            + C_{k^2\omega^4}^{\mathrm{cw}(n=0,1)} k^2
            + C_{k^4\omega^4}^{\mathrm{cw}(n=0,1)} k^4
            ) (\hbar\omega)^4
         + ( C_{k^1\omega^3}^{\mathrm{cw}(n=0,1)} k 
            + C_{k^3\omega^3}^{\mathrm{cw}(n=0,1)} k^3
            ) (\hbar\omega)^3  \nonumber \\
    & + & ( C_{k^2\omega^2}^{\mathrm{cw}(n=0,1)} k^2 
             + C_{k^4\omega^2}^{\mathrm{cw}(n=0,1)} k^4 
             + C_{k^6\omega^2}^{\mathrm{cw}(n=0,1)} k^6 
             + C_{k^8\omega^2}^{\mathrm{cw}(n=0,1)} k^8 
             ) (\hbar\omega)^2   \nonumber \\
    & + & ( C_{k^3\omega}^{\mathrm{cw}(n=0,1)} k^3 
              + C_{k^5\omega}^{\mathrm{cw}(n=0,1)} k^5 
              + C_{k^7\omega}^{\mathrm{cw}(n=0,1)} k^7 
              ) (\hbar\omega)  \nonumber \\
    & + & C_{k^4\omega^0}^{\mathrm{cw}(n=0,1)} k^4 
             + C_{k^6\omega^0}^{\mathrm{cw}(n=0,1)} k^6 
             + C_{k^8\omega^0}^{\mathrm{cw}(n=0,1)} k^8 
             + C_{k^{10}\omega^0}^{\mathrm{cw}(n=0,1)} k^{10} 
             + C_{k^{12}\omega^0}^{\mathrm{cw}(n=0,1)} k^{12} .
\end{eqnarray}
There are six dispersion curves, corresponding, at any given
waveumber, to six mutually orthogonal phonon perturbations.
The coefficients $C_{k^i \omega^j}^{\mathrm{cw}(n=0,1)} = (i!
j!)^{-1} \partial_k^i \partial_\omega^j |M|$ are too lengthy to be
included explicitly in full in the body of the text.
From the band diagrams, $\omega=\omega_j(k)$, come the energies of the
phonons, $\hbar \mathrm{Re}[\omega_j(k)]$, the MI
$\mathrm{Im}[\omega_j(k)]$, phase velocities $v_p =
\mathrm{Re}[\omega_j(k)] / k$, group velocities $v_g = (d/dk) \{
\mathrm{Re}[\omega_j(k)] \}$, phonon group velocity dispersion values
$(d^2/dk^2) \{ \mathrm{Re}[\omega_j] \} = (d/dk) v_g$, and
higher-order dispersion $(d^n/dk^2) \{ \mathrm{Re}[\omega_j] \}$,
where $n>2$.
Solving involves finding the roots of a sixth-order polynomial.
Numerical solutions may be readily obtained.  The general case is not
analytically soluble. 
%
%
There are analytic solutions in some limiting and special cases.

In the case with zero particle density of spin $M_F=0$, the equation for the
phonons in spin states $M_F=\pm 1$ is the same as for the CNLS equation,
\begin{equation} 
\label{Eq.CNLS.phonon_dispersion_curve}
0 = (\hbar\omega)^4
   + ( C_{k^2\omega^2}^\mathrm{CNLS} k^2
      + C_{k^4\omega^2}^\mathrm{CNLS} k^4
      ) (\hbar\omega)^2
   + C_{k^3\omega}^\mathrm{CNLS} k^3 (\hbar\omega)  
   + C_{k^4\omega^0}^\mathrm{CNLS} k^4 
   + C_{k^6\omega^0}^\mathrm{CNLS} k^6 
   + C_{k^8\omega^0}^\mathrm{CNLS} k^8 ,
\end{equation}
which are well-known from optics.  This is in principle soluble,
though the solutions for a quartic polynomial is long even with simple
coefficients, and for spinor BECs the coefficients are not very short. 

\subsection{Special cases with full analytic phonon bands}

There are a few cases for which the whole dispersion band function
is available analytically.
If two of the three spin fields are zero, the frequency of the cw is
$\hbar \omega_j = \frac{\hbar k_j^2}{2 m} + 2 c A_j^2$, where $c = c_0
+ c_2$ for spin $M_F = j = \pm 1$, and $c = c_0$ for spin $M_F = j
=0$.  Sound waves on top of this cw have frequency
\begin{equation}
\hbar \omega =
\frac{\hbar^2 k_j}{m} k
\pm
\sqrt{\frac{\hbar^2 k^2}{2 m} 
\left( \frac{\hbar^2 k^2}{2 m} + 2 c A_j^2 \right)} .
\label{Bogoliubov_dispersion}
\end{equation}
Equation~(\ref{Bogoliubov_dispersion}) is the well-known Bogoliubov
dispersion relation \cite{Bogolubov.1947}.  When the nonlinearity is
repulsive or zero ($c \geq 0$), the frequencies are real-valued, and
all the cw solutions are stable.  When the nonlinearity is attractive
($c < 0$), there are sound waves with complex-valued frequencies,
i.e., MI, for wave numbers $\hbar^2 k^2 / (2 m) < -2 c A_j^2$.  The
largest MI is at wave numbers $k = \pm \sqrt{-2 m c} A_j / \hbar$, and
the growth rate is $\mathrm{Im}(\omega) = -c A_j^2 / \hbar$
\cite{BespalovTalanov.1966}.

For perturbations on top of cws with nil $M_F=0$ fields, the
eigenvalues (energies, frequencies, chemical potentials) are roots of
fourth-order polynomials, which can be solved in terms of roots.  In
the limiting case in which the wave numbers are the same
($k_1=k_{-1}$), the phonon frequency as a function of wave number,
i.e., the band diagram, is
\begin{equation}
\hbar \omega =
q B^2
\pm \left( \frac{\hbar^2 k^2}{2m} \right)^{1/2}
\left(
  \frac{\hbar^2 k^2}{2m}
+ (c_0+c_2) (A_1^2 + A_{-1}^2)
\pm \left\{
         \left[
           (c_0+c_2) (A_1^2 + A_{-1}^2) 
        \right]^2 
      - 16 c_0 c_2 A_1^2 A_{-1}^2 
      \right\}^{1/2}
\right)^{1/2} .
\label{CNLS.dk0.E_vs_k}
\end{equation}
Two of the four branches of the phonon solutions in
Eq.~(\ref{CNLS.dk0.E_vs_k}) contains frequencies with complex parts,
i.e., modulational instabilities, when the BEC is ferromagnetic,
$c_2<0$, such as is the case for $^{87}$Rb; the frequencies are
real-valued for anti-ferromagnetic BECs, $c_2 \geq 0$.  There are
analytic solutions for phonons in the general CNLS case, where the cw
contains non-zero wave number differences.  However, we found the
general formulas for sound waves in the $M_F=\pm 1$ fields to be long and
complex to the point that computational efficiency but little
understanding could be gained from the exact solutions.

For the ferromagnetic CNLS--type phonons, the largest MI (imaginary
component of the frequency) and the wave numbers (up to a sign, and
offset with respect to the average of the wave numbers) at which they
occur, $(k_1-k_{-1})/2$, are
\begin{subequations}
\begin{eqnarray}
\omega_\mathrm{MI}^\mathrm{CNLS} (k_1=k_{-1})
& \equiv & \frac{1}{2 \hbar} \left[ \sqrt{ [(c_0+c_2) (A_1^2 + A_{-1}^2)]^2 
- 16 c_0 c_2 A_1^2 A_{-1}^2 } - (c_0+c_2) (A_1^2 + A_{-1}^2) \right] , \\
k_\mathrm{MI}^\mathrm{CNLS}(k_1=k_{-1})
& \equiv & \sqrt{ \frac{2 m}{\hbar} \omega_\mathrm{MI}^\mathrm{CNLS} 
(k_1=k_{-1}) }  ,
\end{eqnarray}
\end{subequations}
which is similar to Bogoliubov dispersion \cite{Bogolubov.1947} and
MI in the NLS equation \cite{BespalovTalanov.1966},
though with different parameters.  The MI range is $|k| < 2
k_\mathrm{MI}^\mathrm{CNLS}(k_1=k_{-1})$.  Since there is no MI above
a certain wavelength, the cw should fail to manifest MI if the domain is periodic with
length less than or equal to $\pi / k_\mathrm{MI}^\mathrm{CNLS} (k_1=k_{-1})$.

For the CNLS case, the small-amplitude (linearized)
$M_F=0$ fields decouple from the $M_F=\pm 1$ phonons, and have simple
analytic solutions,
\begin{equation}
\hbar \omega = \pm \left\{ \left[ \frac{\hbar^2 k^2}{2 m} + c_2 (A_1^2
+ A_{-1}^2) - q B^2 - \frac{\hbar^2}{2 m} \left( \frac{k_1 - k_{-1}}{2} \right)^2 \right]^2 - 4 c_2^2 A_1^2 A_{-1}^2 \right\}^{1/2} \, .
\end{equation}
The cw will be modulationally stable in the $M_F=0$ field if
\begin{equation}
E_\mathrm{MI}^\mathrm{crit} 
\equiv
c_2 [A_1 - \mathrm{sign}(c_2) A_{-1}]^2 - q B^2 
- \frac{\hbar^2}{2 m} \left( \frac{k_1 - k_{-1}}{2} \right)^2 
\geq
0 ,
\end{equation} 
and subject to MI if $E_\mathrm{MI}^\mathrm{crit} < 0$.  In the
unstable case, the maximum MI values occur at wave number (recall that
the system is being analyzed in a reference frame in which $k_1+k_{-1}=0$)
$k=0$ if
\begin{equation}
c_2 (A_1^2 + A_{-1}^2) - q B^2 - (\hbar^2/2 m) [(k_1 - k_{-1})/2]^2 \geq 0 ,
\end{equation}
yielding peak MI
\begin{equation}
\max[\mathrm{Im}(\omega)] = \hbar^{-1} \sqrt{4 c_2^2 A_1^2 A_{-1}^2 - [c_2 (A_1^2 + A_{-1}^2) - q B^2 - \hbar^2 (k_1-k_{-1})^2 / (8 m)]^2} \, ;
\end{equation}
otherwise, the maximum MI occurs at phonons that are combinations of
\begin{subequations}
\begin{equation}
k = \pm \sqrt{(k_1 - k_{-1})^2 / 4 - (2 m / \hbar^2) [c_2 (A_1^2 + A_{-1}^2) - q B^2]} ,
\end{equation}
yielding peak MI
\begin{equation}
\max[\mathrm{Im}(\omega)] = 2 |c_2| A_1 A_{-1} / \hbar .
\end{equation}
\end{subequations}

There are analytic solutions for the band diagram of sound waves on
$n=0$ type cws in the limiting case in which the wave numbers are all
the same ($k_1=k_0=k_{-1}$) and the quadratic Zeeman splitting is
zero,
\begin{subequations}
\label{SpinorBEC.dk0.n0.E_vs_k}
\begin{eqnarray}
\omega & = & \pm \frac{\hbar k^2}{2 m} , \\
\omega & = & \pm \frac{1}{\hbar} \left[ \frac{(\hbar k)^2}{2 m} - 
2 c_2 (A_1+A_{-1})^2  \right] , \\
\omega & = & \pm \frac{1}{\hbar} \sqrt{ \frac{(\hbar k)^2}{2 m} 
\left[ \frac{(\hbar k)^2}{2 m} + 2 (c_0+c_2) (A_1+A_{-1})^2 \right] } .
\end{eqnarray}
\end{subequations}
It follows from Eqs.~(\ref{SpinorBEC.dk0.n0.E_vs_k}) that the cws have
MI if and only if $c_0+c_2<0$.  This is not the case for either
$^{23}$Na or $^{87}$Rb, so for these the $n=0$ cws are stable when the
wave numbers are all the same and quadratic Zeeman splitting is
absent.  Thus, ironically, in ferromagnetic BECs, the ($n=0$)--type cw
solutions are modulationally stable and the CNLS--type solutions have
MI, even though the former cws have higher energy than the latter.  We
failed to find analytic formulas for the phonons when there was
quadratic Zeeman splitting or non-zero differences in the wave numbers
of the spin components $M_F$.

\subsection{Small and large wave number (\textit{k}) limiting cases}

\subsubsection{Continuous wave background solutions of types \textit{n}=0,1}

In the large wave number limit, the equation for the band diagram for
sound waves on top of the cw($n=0,1$) solutions approaches (in
dimensionless units)
\begin{eqnarray}
0 & = & (\omega^2/k^4)^3 
         + C_{k^4\omega^4}^{\mathrm{cw}(n=0,1)} (\omega^2/k^4)^4 
         + C_{k^8\omega^2}^{\mathrm{cw}(n=0,1)} (\omega^2/k^4) 
         + C_{k^{12}\omega^0}^{\mathrm{cw}(n=0,1)} \nonumber \\
   & = & (\omega^2/k^4)^3 - (3/4) (\omega^2/k^4)^4 + (3/16) (\omega^2/k^4) 
   - 1/64  \nonumber \\
   & = & (\omega^2/k^4 - 1/4)^3 ,
\end{eqnarray}
or, with the dimensions left in, $\omega = \pm \hbar k^2/(2 m)$.  At
large wave numbers, the kinetic terms dominate over the
nonlinearities, and the dispersion approaches a quadratic dependence
on the wave number (i.e., constant dispersion), the same as it would be
in the absence of nonlinearities.
In the limit of small wave numbers ($k\approx 0$), the dispersion
curves can be obtained by substituting a Taylor expansion,
\begin{equation}
\omega(k) = \omega_0 +  \omega_1 k +  \frac{1}{2} \omega_2 k^2 + \ldots  \, ,
\end{equation}
into the equation~(\ref{Eq.phonon_dispersion_curve}) for the complete
dispersion curve.

Two of the phonon dispersion curves at low momentum (small $|k|$) have
Taylor coefficients
\begin{subequations}
\begin{eqnarray}
\omega_0 & = & \pm \sqrt{-C_{k^0\omega^4}^{\mathrm{cw}(n=0,1)}} , \\ 
\omega_1 & = & C_{k^1\omega^3}^{\mathrm{cw}(n=0,1)} / 
(2 C_{k^0\omega^4}^{\mathrm{cw}(n=0,1)}) , \\
\omega_2 & = & -\frac{1}{\omega_0} 
                        \left( 
                          C_{k^2\omega^4}^{\mathrm{cw}(n=0,1)} 
                        - \frac{C_{k^2\omega^2}^{\mathrm{cw}(n=0,1)}}
			{C_{k^2\omega^4}^{\mathrm{cw}(n=0,1)}} 
                        +  \frac{3 C_{k^1\omega^3}^{\mathrm{cw}(n=0,1)}}
			{4 (C_{k^0\omega^4}^{\mathrm{cw}(n=0,1)})^2}
                        \right) ,
\end{eqnarray}
\end{subequations}
and four dispersion curves have $\omega_0=0$ for the 0th-order in the
expansion, and the linear terms in the dispersion curves are the roots
of the quartic equation in $\omega_1$,
\begin{equation} 
0 = C_{k^0\omega^4}^{\mathrm{cw}(n=0,1)} \hbar^4 \omega_1^4
   + C_{k^1\omega^3}^{\mathrm{cw}(n=0,1)} \hbar^3 \omega_1^3
   + C_{k^2\omega^2}^{\mathrm{cw}(n=0,1)} \hbar^2 \omega_1^2  
   + C_{k^3\omega}^{\mathrm{cw}(n=0,1)} \hbar \omega_1 
   + C_{k^4\omega^0}^{\mathrm{cw}(n=0,1)} .
\end{equation}
Explicit analytic formulas for higher orders in the expansion of the
dispersion curves do not seem very useful, so we stop at this order.

\subsubsection{cw background solutions with zero particle density for  spin \textit{m}=0}

In the large wave number limit, the band diagram of phonons on top of
cw solutions without an $M_F=0$ component approaches $\omega = \pm
\hbar k^2/(2 m)$.  At large wave numbers, the kinetic terms dominate
over the nonlinearities, and the dispersion approaches a quadratic
dependence on the wave number (i.e., constant dispersion), the same as
it would be in the absence of nonlinearities.
In the limit of small wave numbers ($k\approx 0$), the dispersion
curves can be obtained by substituting a Taylor expansion
\begin{equation}
\omega(k) = \omega_0 +  \omega_1 k +  \frac{1}{2} \omega_2 k^2 + .... 
\end{equation}
into the equation~(\ref{Eq.CNLS.phonon_dispersion_curve}) for the CNLS
dispersion curve.
The 0th-order term is nil ($\omega_0$).  The first-order terms
($\omega_1$)---one for each of the four curves---are the solutions of
the quartic polynomial
\begin{equation} 
0 = \hbar^4 \omega_1^4
   + C_{k^2\omega^2}^\mathrm{CNLS} \hbar^2 \omega_1^2
   + C_{k^3\omega}^\mathrm{CNLS} \hbar \omega_1
   + C_{k^4\omega^0}^\mathrm{CNLS} .
\end{equation}
We stop at the linear expansion terms because general explicit
analytic formulas for the higher-order terms in the series are not
very helpful.  One may carry out the expansion numerically for
specific cw solutions.

\subsection{Band diagram illustrations}

\begin{figure}[hbt]
\centering
\includegraphics[width=8.5cm]{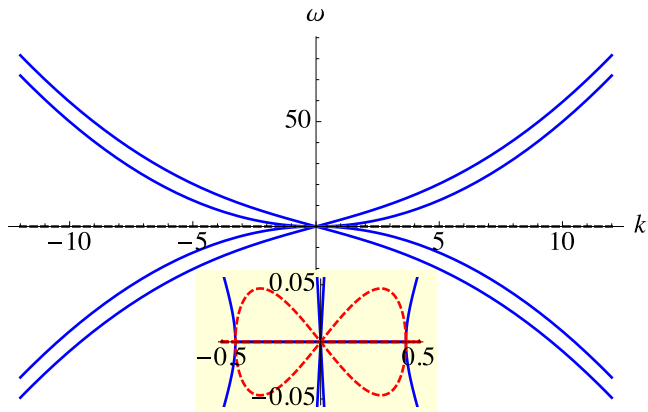}
\caption{(Color online) Band diagram (frequency as a function of wave
vector) for phonons propagating on top of a CNLS--type cw solution
(vanishing $M_F=0$ field) of a $^{87}$Rb BEC where the $M_F=1$ and
$M_F=-1$ spin components have the same wave vectors, and 
amplitudes $2$ and $2.5$.  The energies (chemical potentials) come
from the real parts of the frequencies ($E=\hbar
\mathrm{Re}[\omega]$), which are shown as solid lines; and the MI
comes from the imaginary parts of the frequencies, which are dotted lines.
All quantities in the figure are dimensionless; see Eqs.~(\ref{nondimensionalization}).}
\label{Fig.sound.Rb87.k0.cnls}
\end{figure}

\begin{figure}[hbt]
\centering
\includegraphics[width=8.5cm]{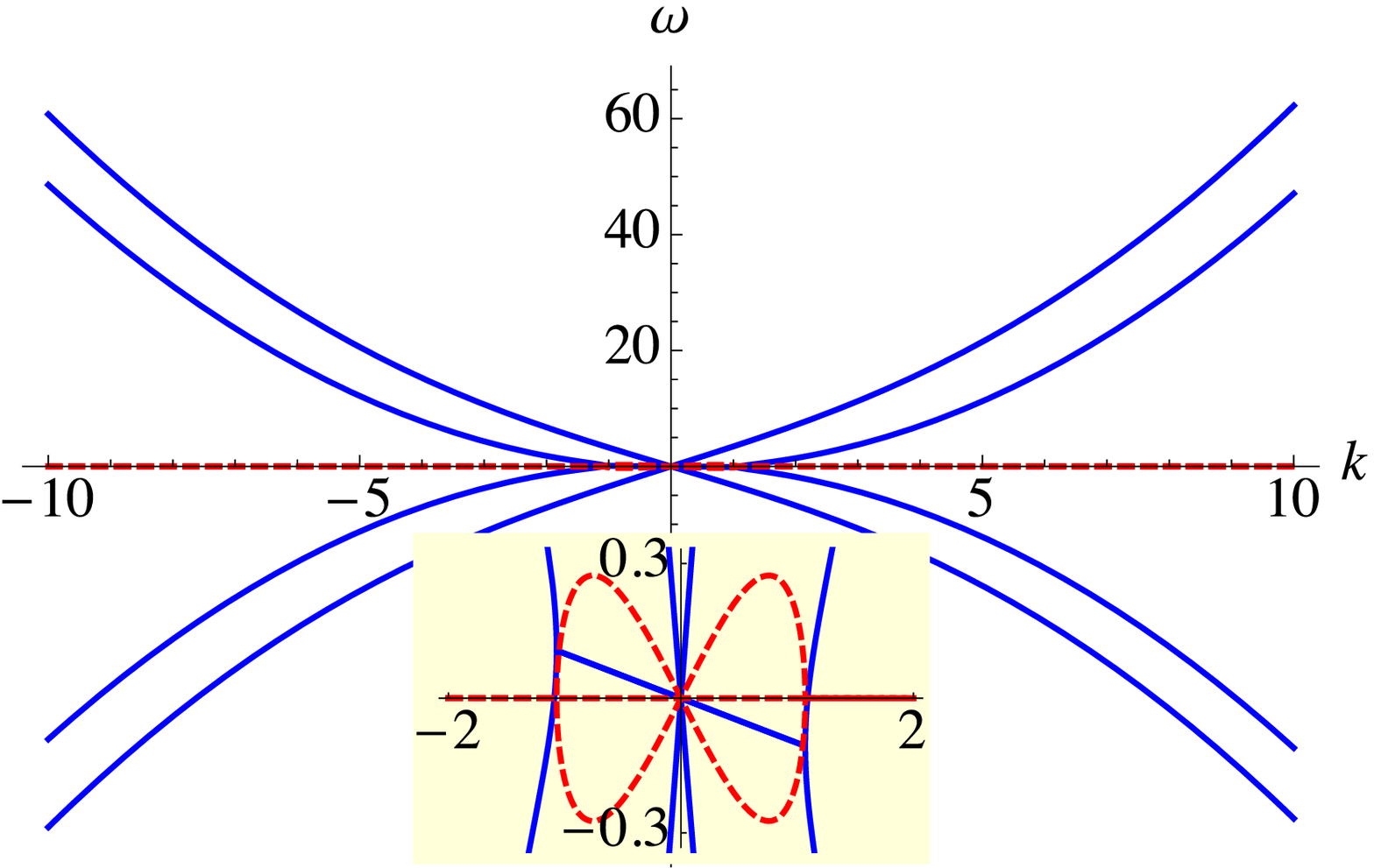}
\caption{(Color online) Band diagram (frequency as a function of wave
vector) for phonons propagating on top of a CNLS--type cw solution
(vanishing $M_F=0$ field) of a $^{87}$Rb BEC where the $M_F = 1$ and
$M_F = -1$ components have
wave vectors that differ by $k_1 - k_{-1} = 1$, and amplitudes $2$ and
$2.5$.  The energies (chemical potentials) come from the real parts of
the frequencies ($E=\hbar \mathrm{Re}[\omega]$), which are shown as
solid lines; and the MI comes from the imaginary parts of the
frequencies, which are dotted lines. 
All quantities in the figure are dimensionless; see Eqs.~(\ref{nondimensionalization}).}
\label{Fig.sound.Rb87.k1.cnls}
\end{figure}

\begin{figure}[hbt]
\centering
\includegraphics[width=8.5cm]{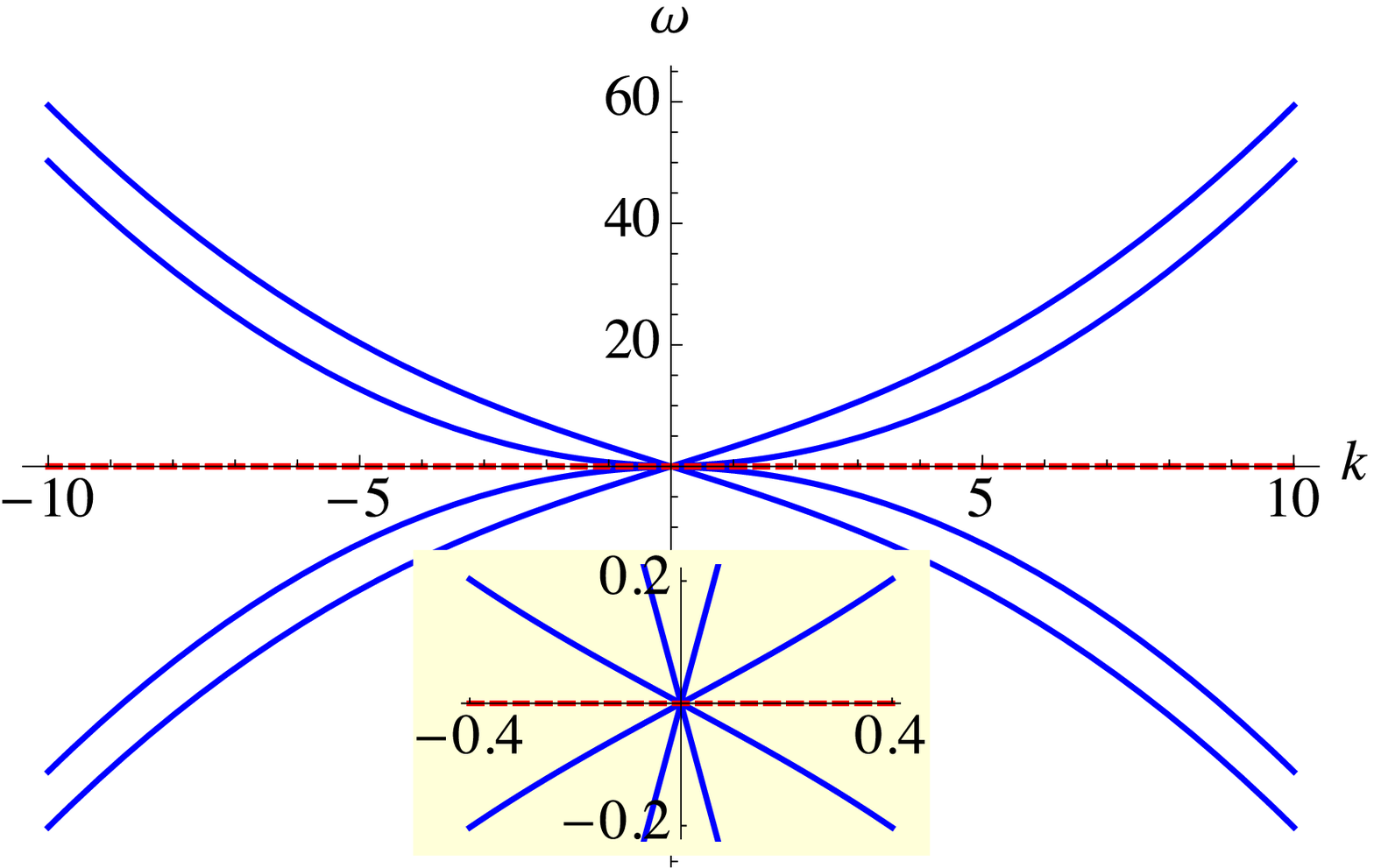}
\caption{(Color online) Band diagram (frequency as a function of wave
vector) for phonons propagating on top of a CNLS--type cw solution
(vanishing $M_F=0$ field) of a $^{23}$Na BEC where the $M_F=1$ and
$M_F=-1$ spin components have the same wave vectors, and 
amplitudes $2$ and $2.5$.  The energies (chemical potentials) come
from the real parts of the frequencies ($E=\hbar
\mathrm{Re}[\omega]$), which are shown as solid lines; and the MI
comes from the imaginary parts of the frequencies, which are dotted
lines.  In this case, the frequencies are real-valued, so there is no MI. 
All quantities in the figure are dimensionless; see Eqs.~(\ref{nondimensionalization}).}
\label{Fig.sound.Na23.k0.cnls}
\end{figure}

\begin{figure}[hbt]
\centering
\includegraphics[width=8.5cm]{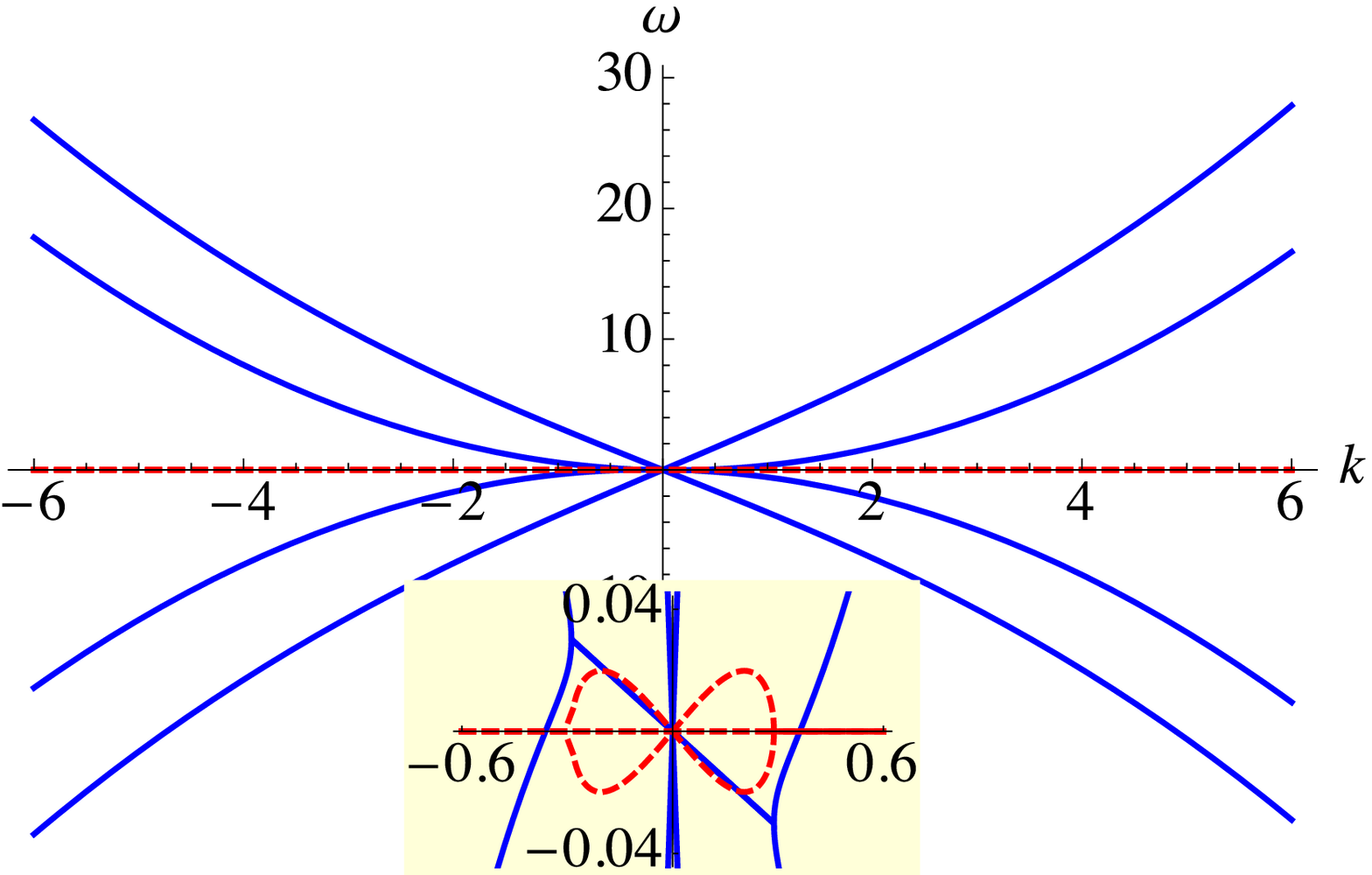}
\caption{(Color online) Band diagram (frequency as a function of wave vector)
for phonons propagating on top of a CNLS--type cw solution
(vanishing $M_F=0$ field) of a $^{23}$Na BEC
where the $M_F = 1$ and $M_F = -1$ components have
wave vectors that differ by $k_1 - k_{-1} = 1$, and amplitudes $2$ and $2.5$. 
The energies (chemical potentials) come from the real parts of the frequencies
($E=\hbar \mathrm{Re}[\omega]$), which are shown as solid lines;
and the MI comes from the imaginary parts of the frequencies,
which are dotted lines.
All quantities in the figure are dimensionless; see Eqs.~(\ref{nondimensionalization}).}
\label{Fig.sound.Na23.k1.cnls}
\end{figure}

\begin{figure}[hbt]
\centering
\includegraphics[width=8.5cm]{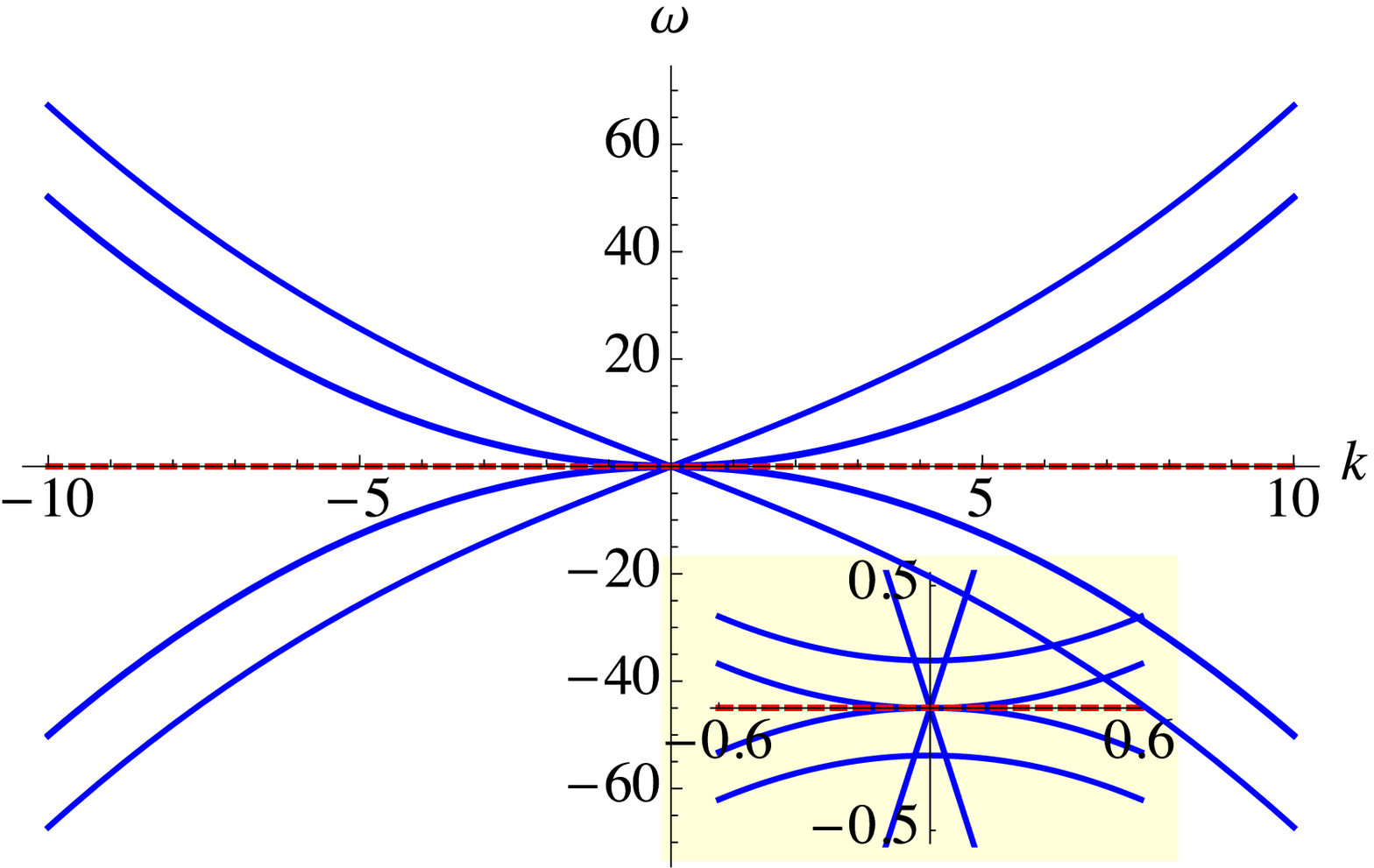}
\caption{(Color online) Band diagram (frequency as a function of wave
vector) for phonons propagating on top of an ($n=0$)--type cw solution of
a $^{87}$Rb BEC where the $M_F=1$ and $M_F=-1$ spin components have
the same wave vectors and amplitudes $2$ and $2.5$.  The
energies (chemical potentials) come from the real parts of the
frequencies ($E=\hbar \mathrm{Re}[\omega]$), which are shown as solid
lines.  The imaginary parts of the frequencies, shown as dotted lines,
are zero, so there is no MI.}
\label{Fig.sound.Rb87.k0.n0}
\end{figure}

\begin{figure}[hbt]
\centering
\includegraphics[width=8.5cm]{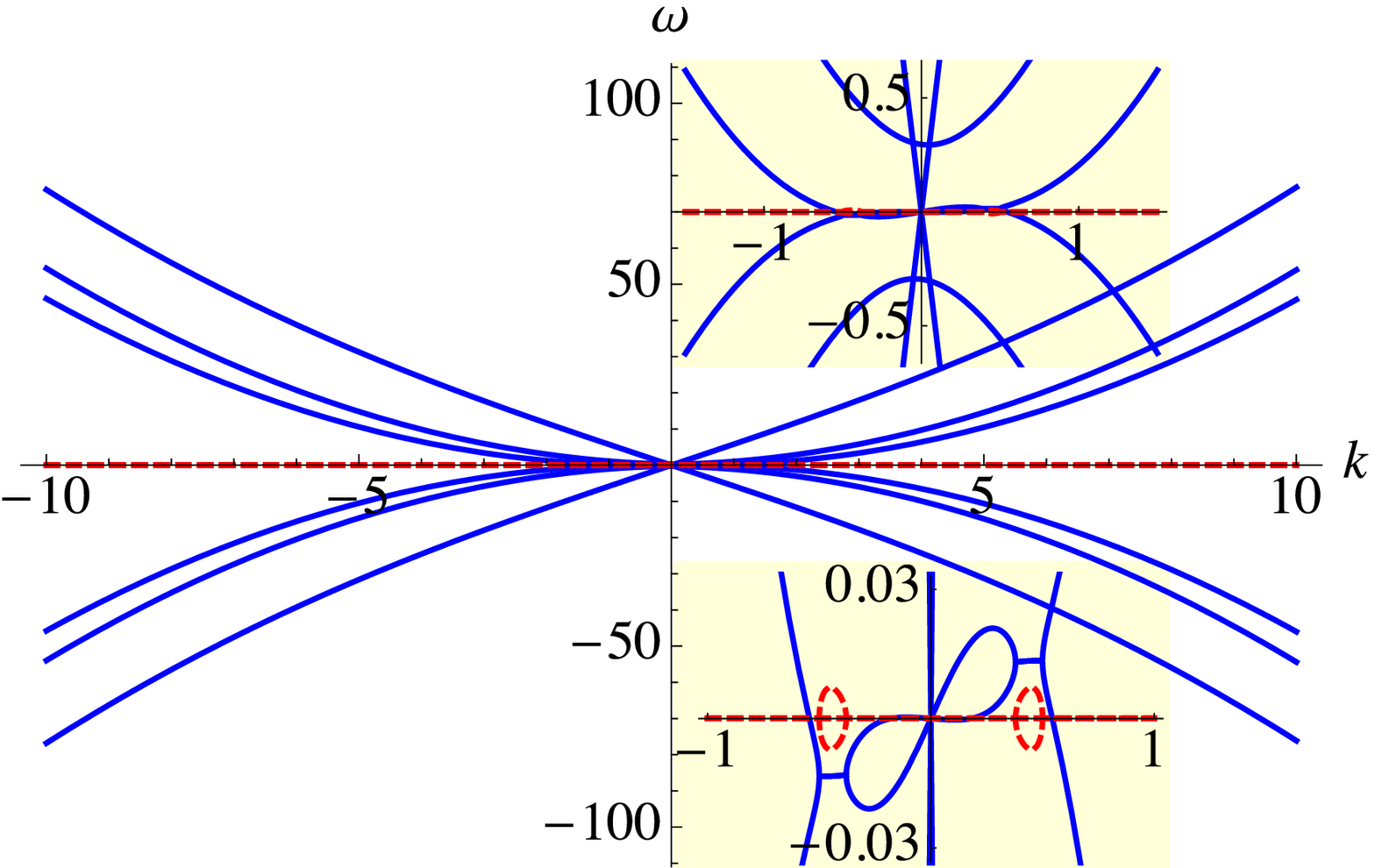}
\caption{(Color online) Band diagram (frequency as a function of wave
vector) for phonons propagating on top of an ($n=0$)--type cw solution of
a $^{87}$Rb BEC where the $M_F = 1$ and $M_F = -1$ components have
wave vectors that differ by $k_1 - k_{-1} = 1$, and amplitudes $2$ and $2.5$. 
The energies (chemical potentials) come from the real parts of the frequencies 
($E=\hbar \mathrm{Re}[\omega]$), which are shown as solid lines; and the MI
comes from the imaginary parts of the frequencies, which are dotted lines.
All quantities in the figure are dimensionless; see Eqs.~(\ref{nondimensionalization}).}
\label{Fig.sound.Rb87.k1.n0}
\end{figure}

\begin{figure}[hbt]
\centering
\includegraphics[width=8.5cm]{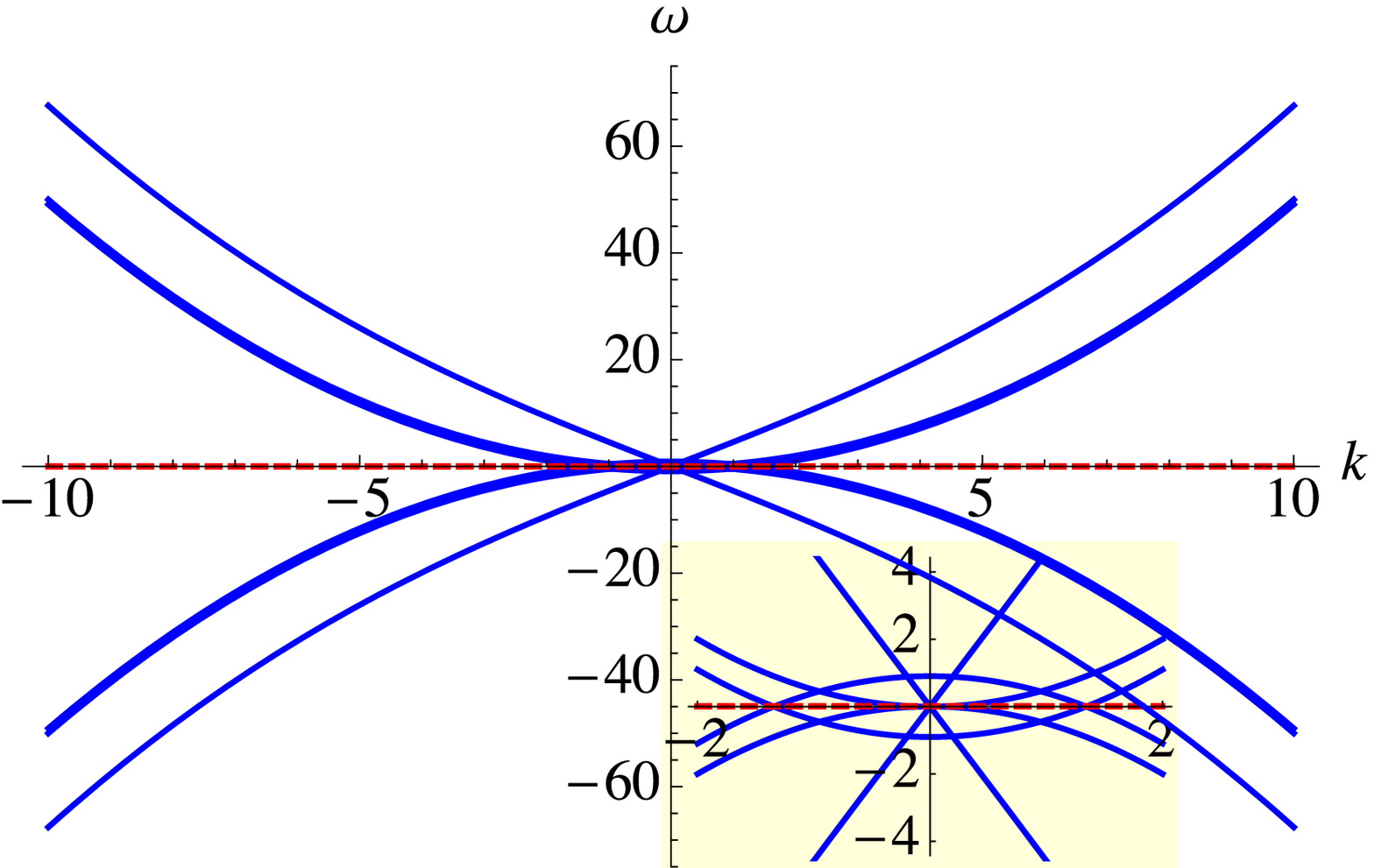}
\caption{(Color online) Band diagram (frequency as a function of wave
vector) for phonons propagating on top of an ($n=0$)--type cw solution of
a $^{23}$Na BEC where the $M_F=1$ and $M_F=-1$ spin components have
the same wave vectors and amplitudes $2$ and $2.5$.  The
energies (chemical potentials) come from the real parts of the
frequencies ($E=\hbar \mathrm{Re}[\omega]$), which are shown as solid
lines.  The imaginary parts of the frequencies, shown as dotted lines,
which would create MI, are zero.
All quantities in the figure are dimensionless; see Eqs.~(\ref{nondimensionalization}).}
\label{Fig.sound.Na23.k0.n0}
\end{figure}

\begin{figure}[hbt]
\centering
\includegraphics[width=8.5cm]{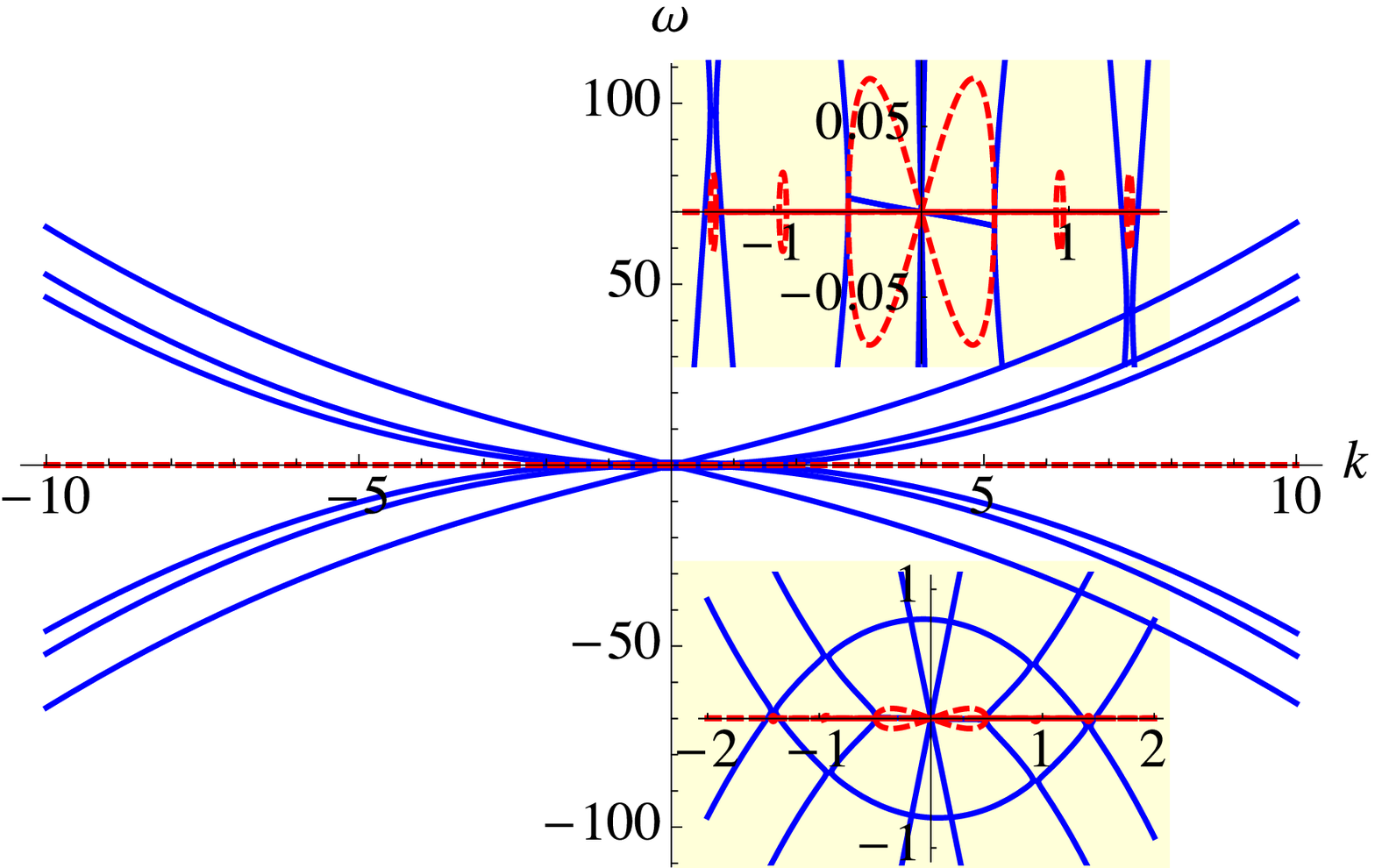}
\caption{(Color online) Band diagram (frequency as a function of wave
vector) for phonons propagating on top of an ($n=0$)--type cw solution of
a $^{23}$Na BEC where the $M_F = 1$ and $M_F = -1$ components have
wave vectors that differ by $k_1 - k_{-1} = 1$,
and amplitudes $2$ and $2.5$.  The energies (chemical
potentials) come from the real parts of the frequencies ($E=\hbar
\mathrm{Re}[\omega]$), which are shown as solid lines; and the MI
comes from the imaginary parts of the frequencies, which are dotted lines.
All quantities in the figure are dimensionless; see Eqs.~(\ref{nondimensionalization}).}
\label{Fig.sound.Na23.k1.n0}
\end{figure}

\begin{figure}[hbt]
\centering
\includegraphics[width=8.5cm]{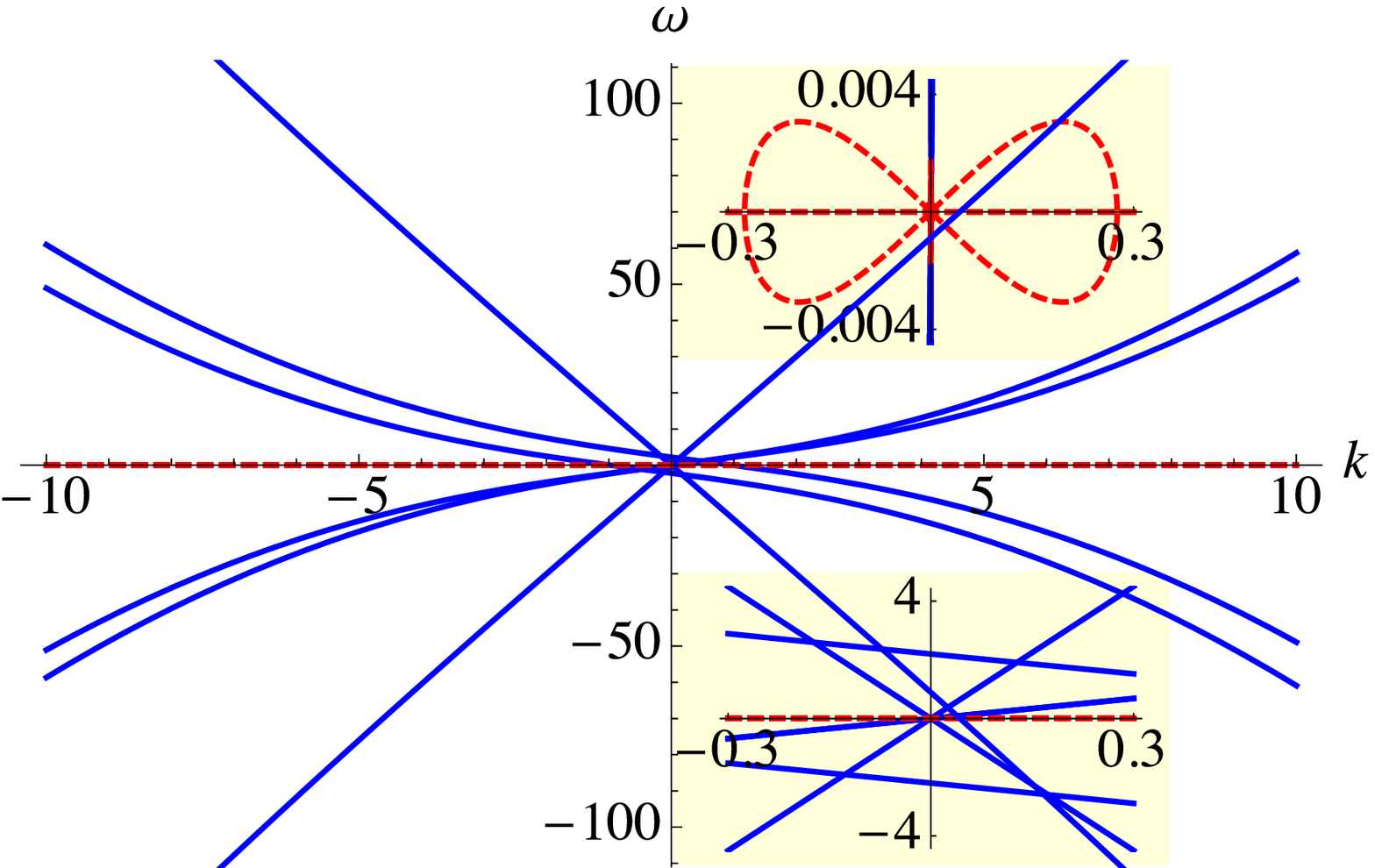}
\caption{(Color online) Band diagram (frequency as a function of wave
vector) for phonons propagating on top of an ($n=1$)--type cw solution of
a $^{23}$Na BEC where the $M_F = 1$ and $M_F = -1$ components have
wave vectors that differ by $k_1 - k_{-1} = 1$,
and amplitudes $2$ and $2.5$.  The energies (chemical
potentials) come from the real parts of the frequencies ($E=\hbar
\mathrm{Re}[\omega]$), which are shown as solid lines; and the MI
comes from the imaginary parts of the frequencies, which are dotted lines.
All quantities in the figure are dimensionless; see Eqs.~(\ref{nondimensionalization}).}
\label{Fig.sound.Na23.k1.n1}
\end{figure}

To illustrate the range of different behaviors in the sound waves and
the MI, we show the real parts of the frequencies
[$\mathrm{Re}(\omega)$] and MI [$\mathrm{Im}(\omega)$] versus
the wave number of sound waves propagating on a background of each of the
allowed cws in a BEC of (i) $^{23}$Na and (ii) $^{87}$Rb, in which the
wave numbers of the different spin components are (a) all identical
and (b) when the wave numbers have different values.  Each of the
examples in
Figs.~\ref{Fig.sound.Rb87.k0.cnls}-\ref{Fig.sound.Na23.k1.n1} takes
the dimensionless amplitudes of the $M_F=\pm 1$ fields to be $A_1=2$
and $A_{-1}=2.5$, and the quadratic Zeeman splitting is zero.

In the CNLS--type solutions, the $M_F=0$ particle density is zero; the
(dimensionless) Hamiltonians for $^{87}$Rb are $52.5191$ when the wave
vectors are the same (corresponding to
Fig.~\ref{Fig.sound.Rb87.k0.cnls}), and $53.8004$ when $k_1-k_{-1}=1$
(corresponding to Fig.~\ref{Fig.sound.Rb87.k1.cnls}); the Hamiltonians
for $^{23}$Na are $52.5877$ when the wave vectors are the same
(corresponding to Fig.~\ref{Fig.sound.Na23.k0.cnls}), and $53.8689$
when $k_1-k_{-1}=1$
(corresponding to Fig.~\ref{Fig.sound.Na23.k1.cnls}).
Of the ($n=0$)--type cw solutions, a $^{87}$Rb BEC has $A_0=3.16228$ and
$H=204.288$ when the wave vectors are the same (corresponding to
Fig.~\ref{Fig.sound.Rb87.k0.n0}), and $A_0=4.78308$ and $H=548.323$
when $k_1-k_{-1}=1$ (corresponding to
Fig.~\ref{Fig.sound.Rb87.k1.n0}); and the ($n=0$)--type cw in a
$^{23}$Na BEC has $A_0=3.16228$ and $H=52.5877$ when the wave vectors
are the same (corresponding to Fig.~\ref{Fig.sound.Na23.k0.n0}), and
$A_0=2.68906$ and $H=156.589$ when $k_1-k_{-1}=1$ (corresponding to
Fig.~\ref{Fig.sound.Na23.k1.n0}).
For the ($n=1$)--type cw solution for $^{23}$Na with $k_1-k_{-1}=1$
(corresponding to Fig.~\ref{Fig.sound.Na23.k1.n1}), $A_0=14.6385$ and
$H=25234.6$.

A sense of the effects of non-zero wave number difference can be gained by comparing
Fig.~\ref{Fig.sound.Rb87.k0.cnls} with Fig.~\ref{Fig.sound.Rb87.k1.cnls}, 
Fig.~\ref{Fig.sound.Na23.k0.cnls} with Fig.~\ref{Fig.sound.Na23.k1.cnls}, 
Fig.~\ref{Fig.sound.Rb87.k0.n0} with Fig.~\ref{Fig.sound.Rb87.k1.n0}, and
Fig.~\ref{Fig.sound.Na23.k0.n0} with Fig.~\ref{Fig.sound.Na23.k1.n0}.
wave number differences in the spin components increase the modulational instability, 
whether the BECs are ferromagnetic or anti-ferromagnetic, with or without a 
nontrivial $M_F=0$ field. Note that the displayed phonon dispersion curve for a
CNLS--type cw in a $^{23}$Na BEC with non-zero difference in the wave numbers (Fig.~\ref{Fig.sound.Na23.k1.cnls})
has zero MI, but similar cws but with smaller particle densities are subject to MI.
This is consistent with the well-known fact that (in the language of optical fibers)
the cw with components of the same wave number in a pair of CNLS equations
without birefringence is stable when the ratio of cross- to self-phase modulation
$B=(c_0-c_2)/(c_0+c_2)$ is less than zero (similar to an anti-ferromagnetic BEC),
and modulationally unstable when the ratio is greater than zero (similar to a ferromagnetic BEC)
\cite{Menyuk.1987,Agrawal.1987,Agrawal.2001}.
This also confirms the known result that---recall that group-velocity birefringence
terms in CNLS equations can be eliminated by a change in variables in which the
frequencies and wave numbers of the cw components are shifted
[see, e.g., Eq.~(7.2.29) in~\cite{Agrawal.2001}]---group-velocity birefringence
(which maps to a difference in the wave numbers of the cw components here)
adds to the MI.
Modulational instability in the CNLS limit with wave number differences has been, for
BECs, referred to as the ``countersuperflow instability.'' \cite{Law.2001, Tsitoura.2013}.
A sense of the difference between ferromagnetic and anti-ferromagnetic BEC can be 
obtained by comparing
Fig.~\ref{Fig.sound.Rb87.k0.cnls} with Fig.~\ref{Fig.sound.Na23.k0.cnls},
Fig.~\ref{Fig.sound.Rb87.k1.cnls} with Fig.~\ref{Fig.sound.Na23.k1.cnls}, 
Fig.~\ref{Fig.sound.Rb87.k0.n0} with Fig.~\ref{Fig.sound.Na23.k0.n0}, and
Fig.~\ref{Fig.sound.Rb87.k1.n0} with Fig.~\ref{Fig.sound.Na23.k1.n0}.
Bose-Einstein condensates of $^{87}$Rb (ferromagnetic) are more subject to MI than
$^{23}$Na (anti-ferromagnetic) for CNLS--type cws and ($n=0$)--type cws
with non-zero wave number difference.  Continuous waves of type $n=0$ with all spin
components at the same wave number are stable against MI for both
$^{87}$Rb and $^{23}$Na.
One of the differences between ferromagnetic and anti-ferromagnetic
BECs is that only the anti-ferromagnetic ones allow ($n=1$)--type cws, such
as with the band diagram in Fig.~\ref{Fig.sound.Na23.k1.n1}.  These
cws have MI, but it is weak compared to MI on the other comparable cw
solutions.
A sense of the differences between the different types of cw solutions
(different values of the amplitudes of the spin $M_F=0$ component for
given $M_F=\pm 1$ fields) may be obtained by comparing
Fig.~\ref{Fig.sound.Rb87.k0.cnls} with Fig.~\ref{Fig.sound.Rb87.k0.n0},
Fig.~\ref{Fig.sound.Rb87.k1.cnls} with Fig.~\ref{Fig.sound.Rb87.k1.n0},
Fig.~\ref{Fig.sound.Na23.k0.cnls} with Fig.~\ref{Fig.sound.Na23.k0.n0},
and Figs.~\ref{Fig.sound.Na23.k1.cnls}, \ref{Fig.sound.Na23.k1.n0}, and~\ref{Fig.sound.Na23.k1.n1}. 
In $^{87}$Rb, the $n=0$ cws are more
stable against MI than the CNLS--type cws.  In $^{23}$Na, there is no
MI in either type of cw when the wave numbers are all the same; when
there is a wave number difference, the $n=1$ cws have the weakest (but
not vanishing) MI, the CNLS cws have stronger MI, and the $n=0$ cws
have the greatest MI.


\subsection{Peak modulational instabilities}

\begin{figure}[hbt]
\centering
\centering\subfigure[]{\includegraphics[width=0.45\textwidth,angle=0]{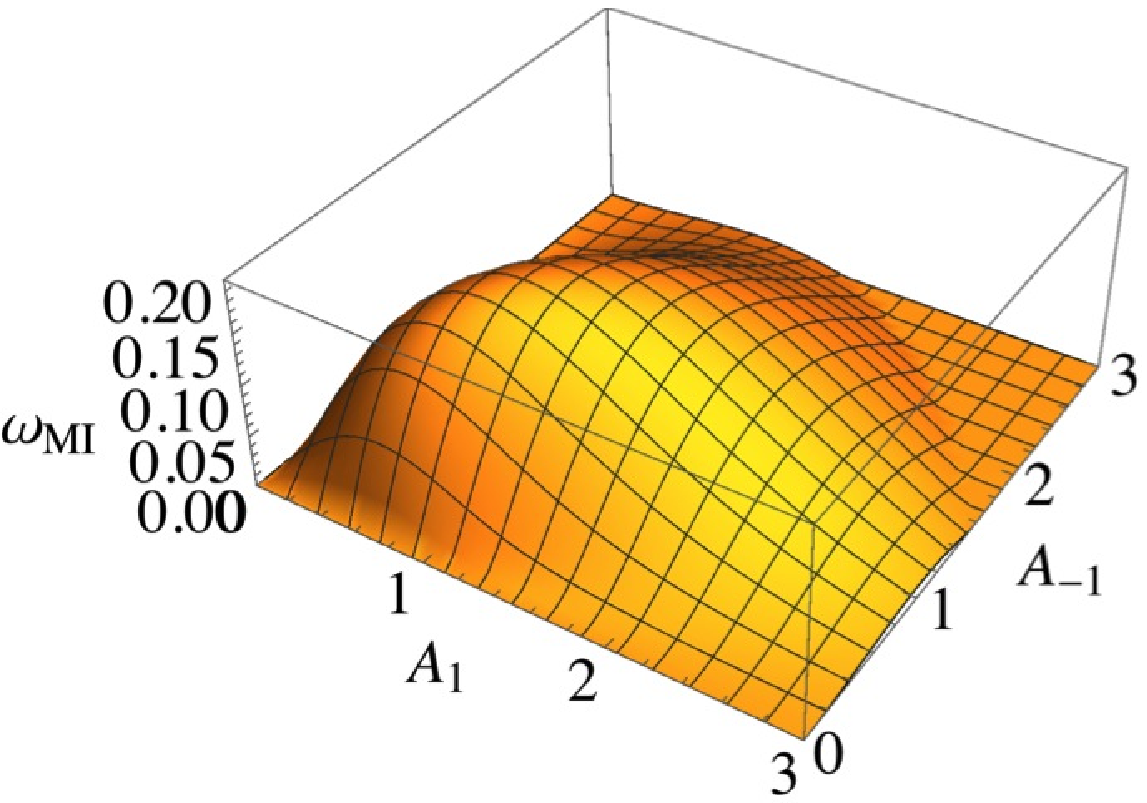}} 
\centering\subfigure[]{\includegraphics[width=0.45\textwidth,angle=0]{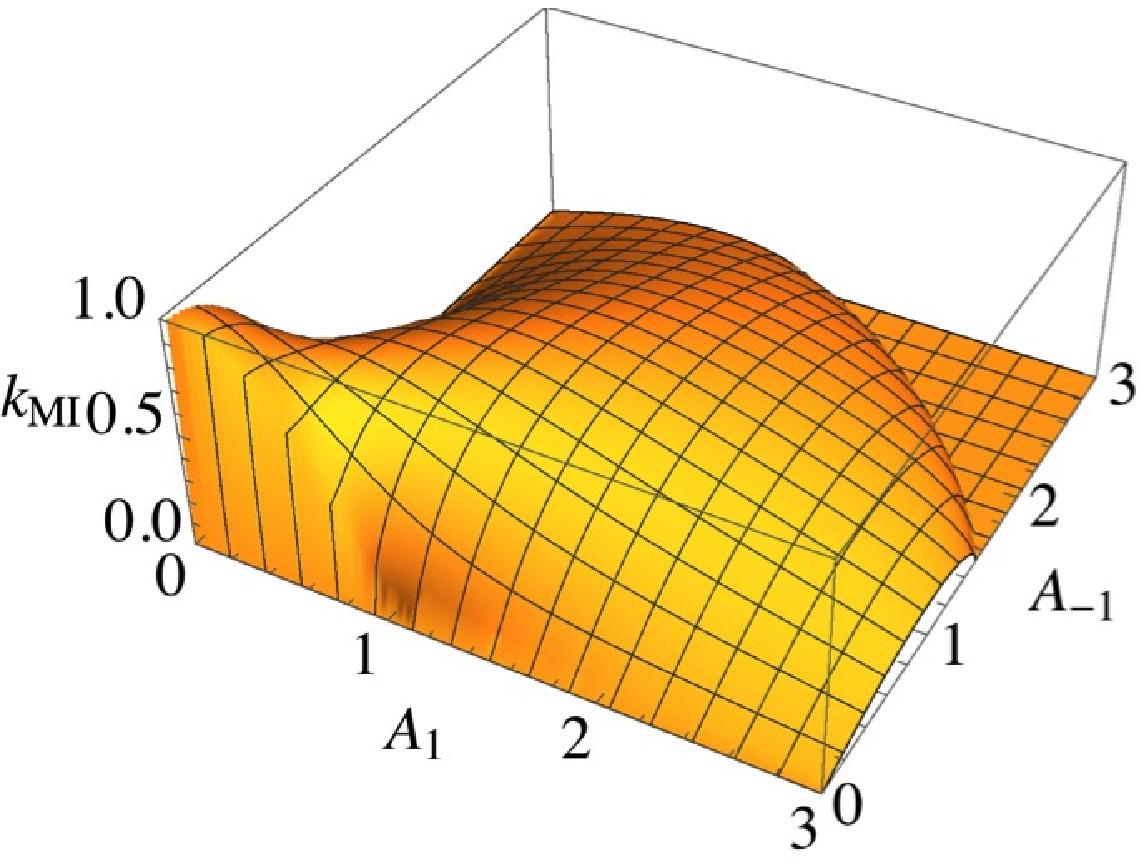}} 
\caption{Maximum MI of CNLS--type (zero $M_F=0$ particle density) cws
of a $^{23}$Na BEC, as a function of the amplitudes of the $M_F=\pm 1$
fields, where the spin components have
wave vectors that differ by $k_1-k_{-1}=1$, and quadratic Zeeman splitting
is zero.  Part~(a) shows the peak MI values, and Part~(b) shows the
wave vectors at which the maxima occur.
All quantities in the figure are dimensionless; see Eqs.~(\ref{nondimensionalization}).}
\label{Fig.Na23.k1.cnls.MI.surf}
\end{figure}

\begin{figure}[hbt]
\centering
\centering\subfigure[]{\includegraphics[width=0.45\textwidth,angle=0]{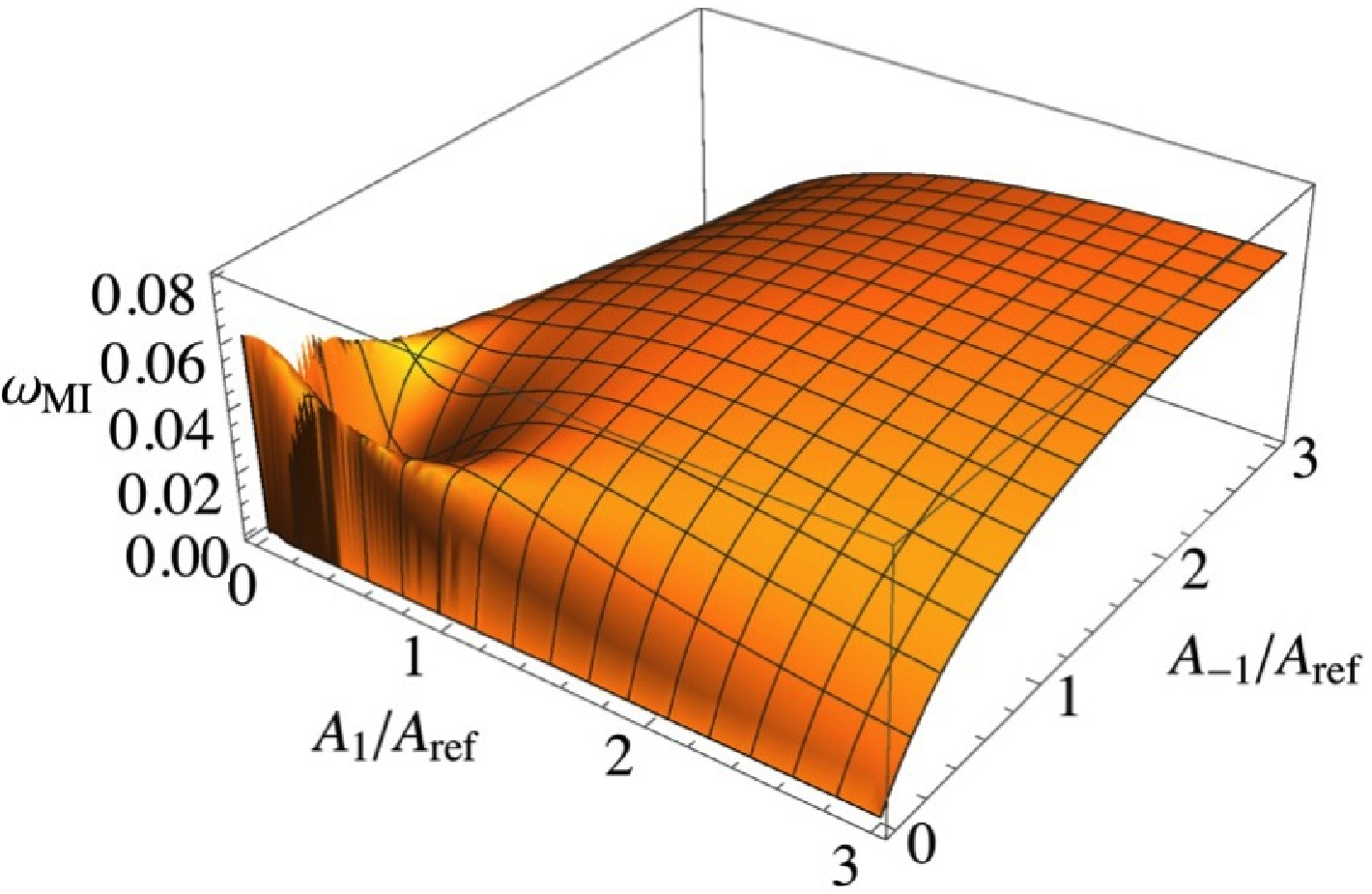}}
\centering\subfigure[]{\includegraphics[width=0.45\textwidth,angle=0]{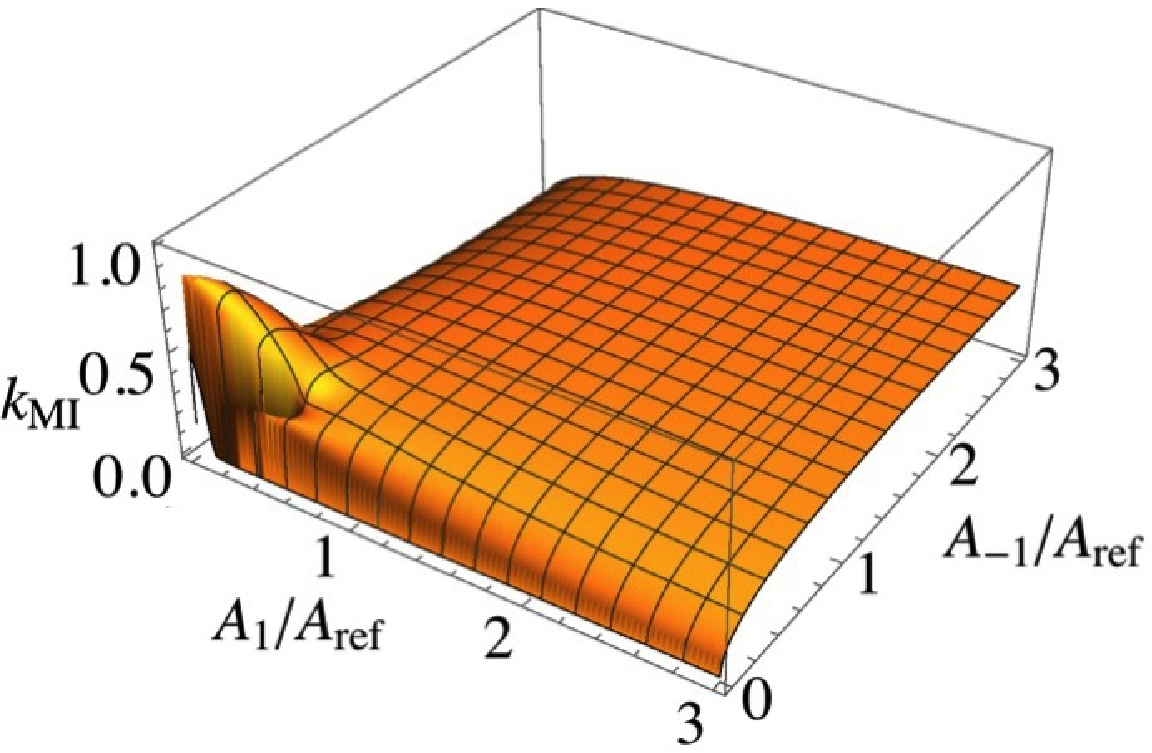}}
\caption{Maximum MI of ($n=0$)--type cws of a $^{87}$Rb BEC, as a
function of the amplitudes of the $M_F=\pm 1$ fields, where the spin
components have wave vectors that differ by $k_1-k_{-1}=1$, 
and quadratic Zeeman splitting is zero.  
The (dimensionless) reference amplitude for this plot is $A_\mathrm{ref}
\equiv \sqrt{| [\hbar^2 (k_1 - k_{-1})^2 / (8m) + q B^2] / c_2|} = (8
|c_2|)^{-1/2} \approx 5.103$.  Part~(a) shows the peak MI values, and
Part~(b) shows the wave vectors at which the maxima occur.
All quantities in the figure are dimensionless; see Eqs.~(\ref{nondimensionalization}).}
\label{Fig.Rb87.k1.n0.MI.surf}
\end{figure}

\begin{figure}[hbt]
\centering
\centering\subfigure[]{\includegraphics[width=0.45\textwidth,angle=0]{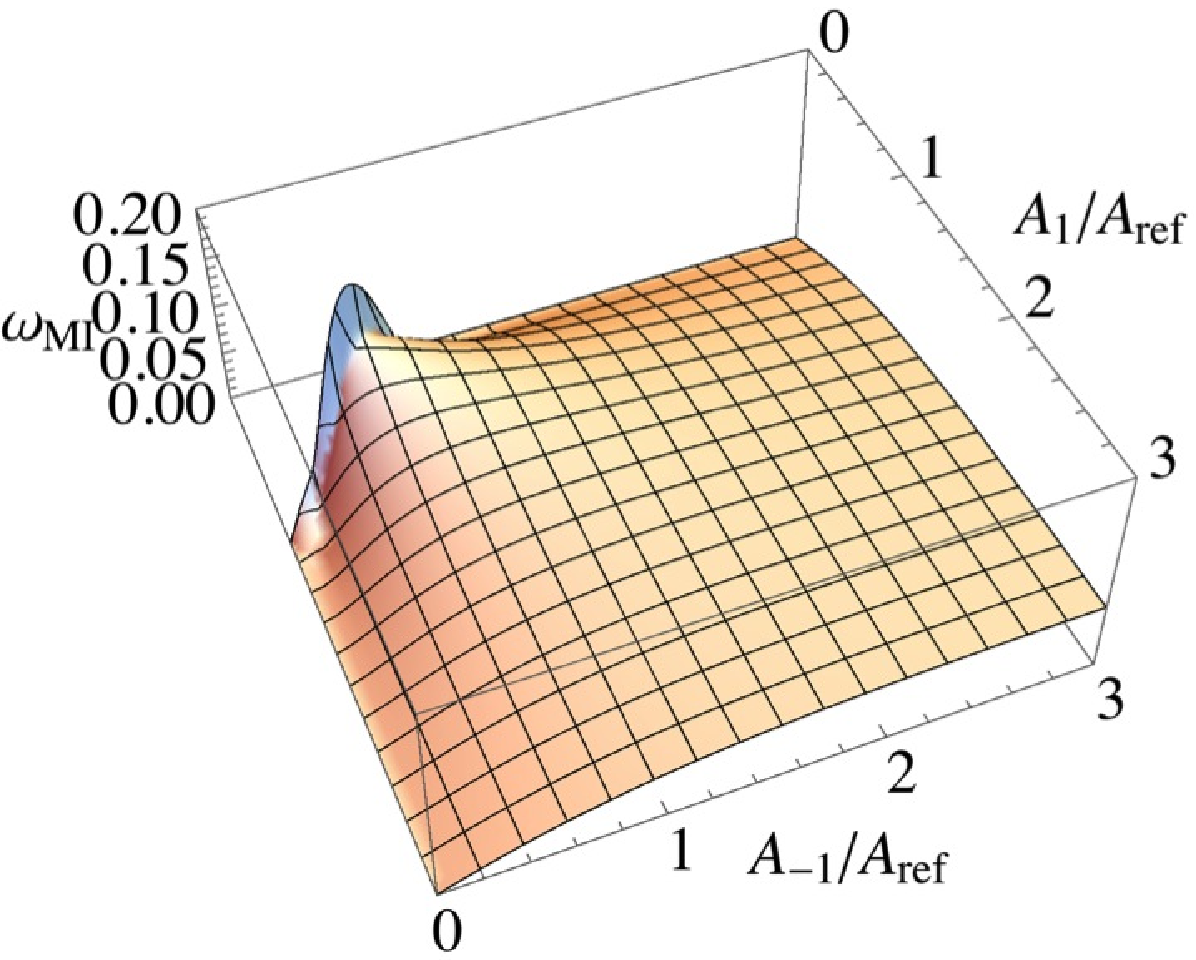}}
\centering\subfigure[]{\includegraphics[width=0.45\textwidth,angle=0]{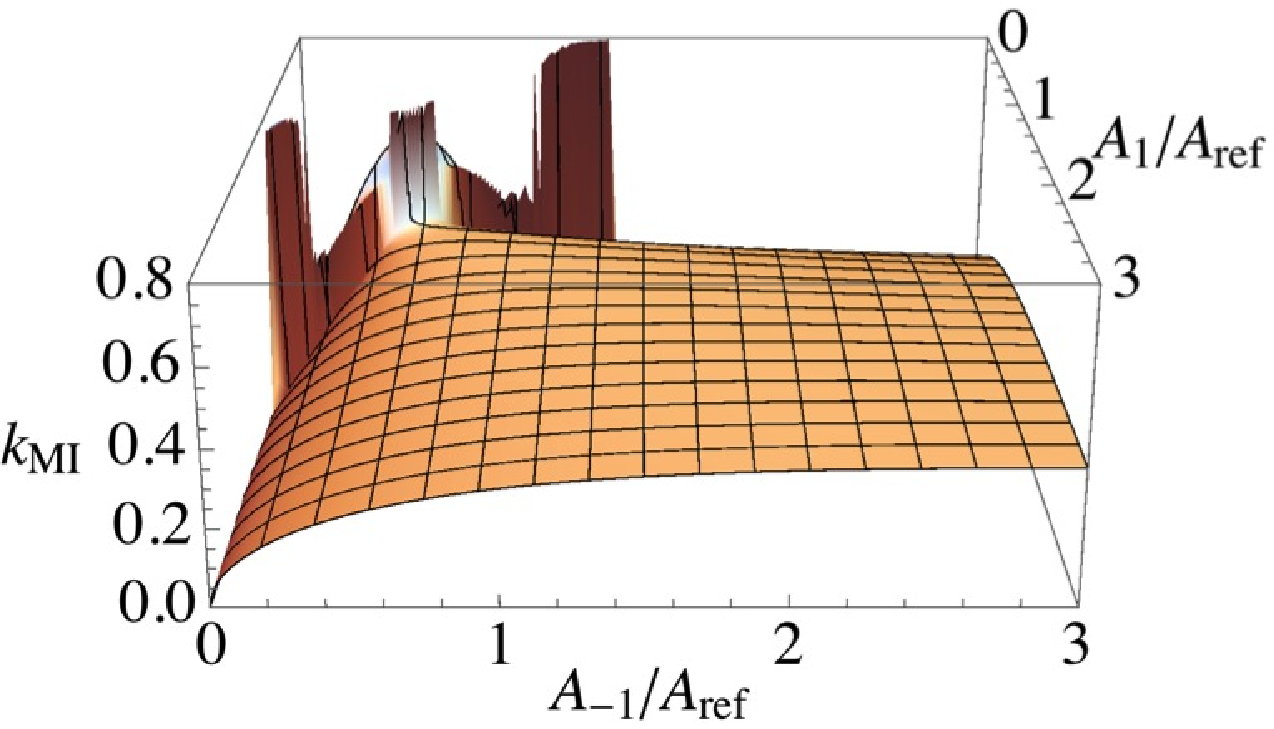}}
\caption{Maximum MI of ($n=0$)--type cws of a $^{23}$Na BEC, as a
function of the amplitudes of the $M_F=\pm 1$ fields, where the spin
components have wave vectors that differ by
$k_1-k_{-1}=1$, and quadratic Zeeman splitting is zero.  The
(dimensionless) reference amplitude for this plot is $A_\mathrm{ref}
\equiv \sqrt{ | [\hbar^2 (k_1 - k_{-1})^2 / (8m) + q B^2] / c_2 |} 
= (8 c_2)^{-1/2} \approx 2.368$.  Part~(a) the peak MI values, and
Part~(b) shows the wave vectors at which the maxima occur.
All quantities in the figure are dimensionless; see Eqs.~(\ref{nondimensionalization}).}
\label{Fig.Na23.k1.n0.MI.surf}
\end{figure}

\begin{figure}[hbt]
\centering
\centering\subfigure[]{\includegraphics[width=0.45\textwidth,angle=0]{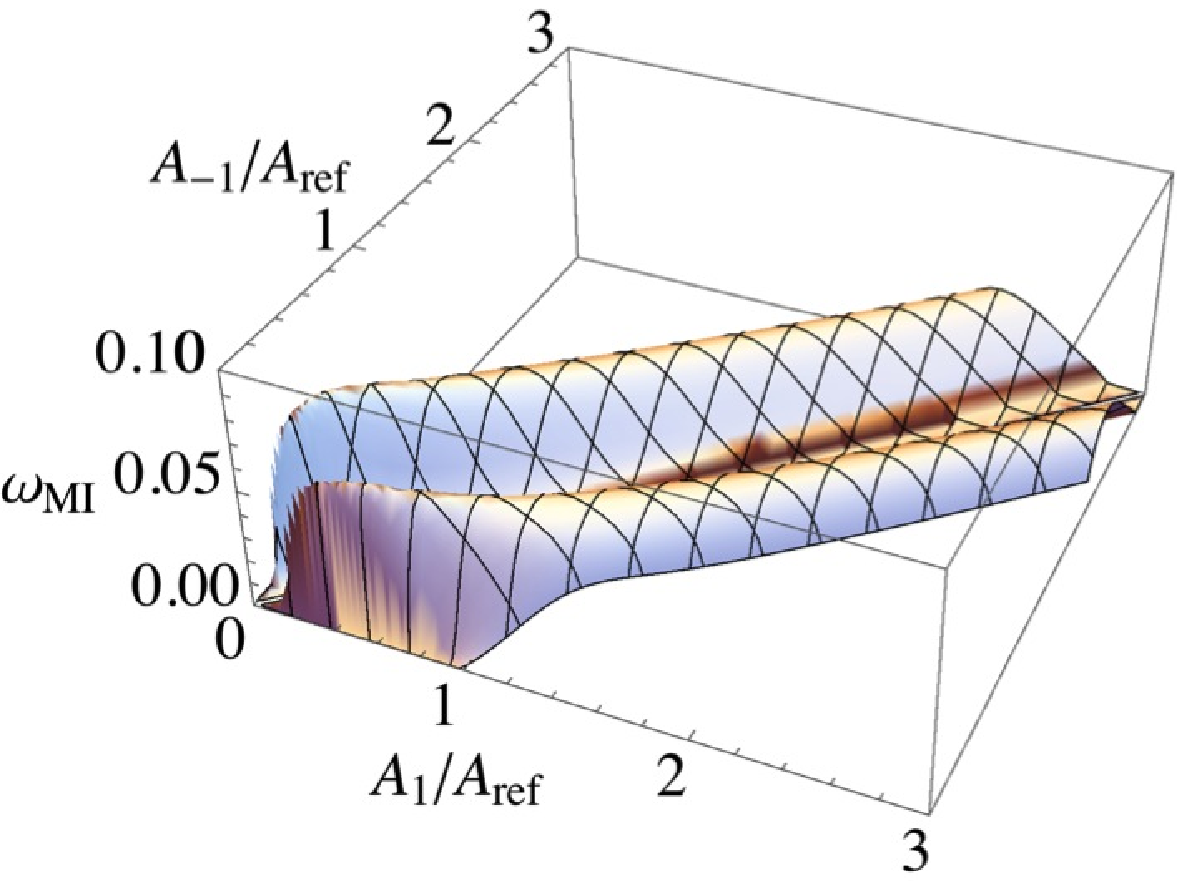}}
\centering\subfigure[]{\includegraphics[width=0.45\textwidth,angle=0]{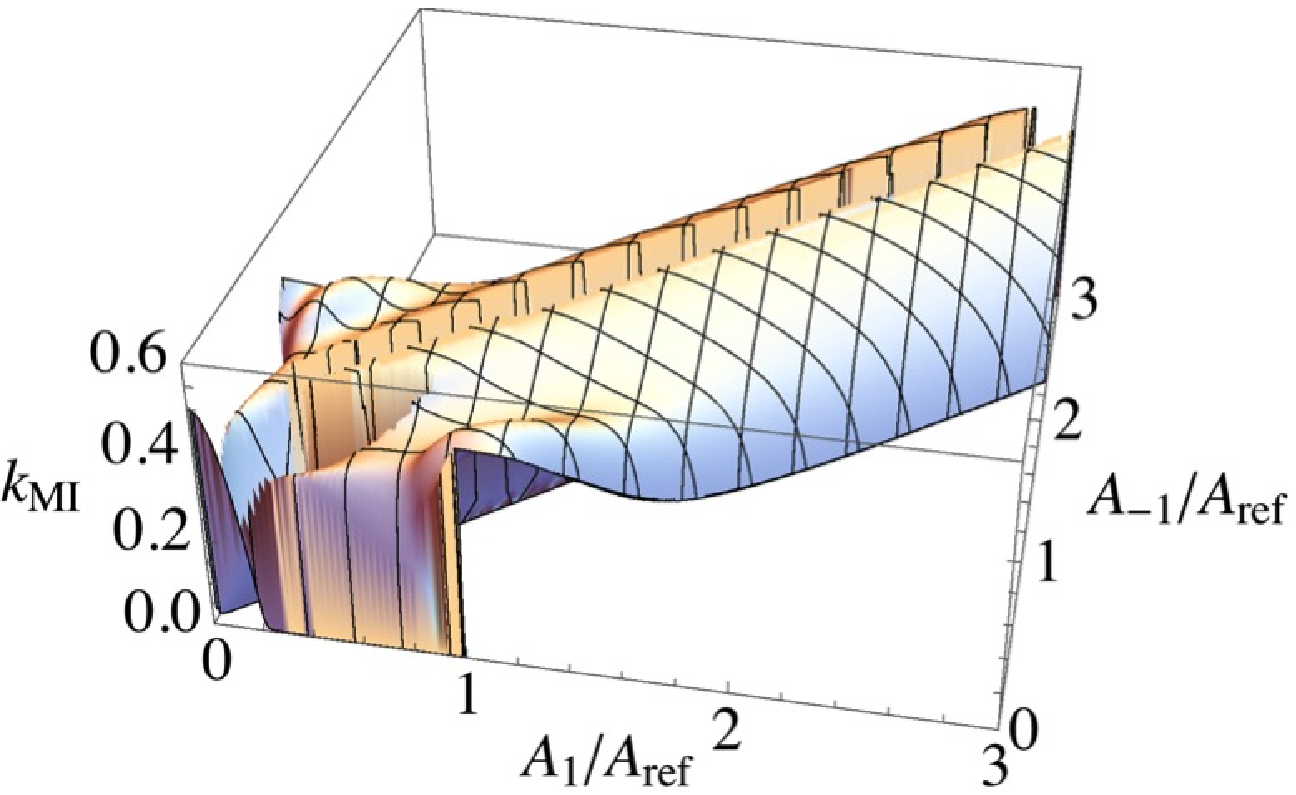}}
\caption{Maximum MI of ($n=1$)--type cws of a $^{23}$Na BEC, as a
function of the amplitudes of the $M_F=\pm 1$ fields, where the spin
components have wave vectors that differ by
$k_1-k_{-1}=1$, and quadratic Zeeman splitting is zero.
The (dimensionless) reference amplitude for this plot is 
$A_\mathrm{ref} \equiv \sqrt{ | [\hbar^2 (k_1 - k_{-1})^2  / (8m) + q B^2] / c_2 |} =  (8 c_2)^{-1/2} \approx 2.368$.
Part~(a) shows the peak MI values, and Part~(b) shows the wave vectors at which the maxima occur.
All quantities in the figure are dimensionless; see Eqs.~(\ref{nondimensionalization}).}
\label{Fig.Na23.k1.n1.MI.surf}
\end{figure}

Modulational instabilities are often more consequential than stable
sound waves because amplification can make them grow from weak to
strong, and the phonons with the largest amplification tend to
dominate after sufficient propagation.  Let us then look at the
maximum MI (the wave number at which the imaginary part of the
frequency is largest, or max$_{j,k} [\omega_j(k)]$, as well as the
value of the wave number $k$ at which the maximum is found).  Since we
are now examining maxima rather than the entire band diagrams, we can
look at larger sections (more dimensions of) the parameter space.  We
plot the maximum MI over \textit{two-dimensional} cross-sections of
the cw parameter space (rather than, as above, for one cw at a time).
Figures~\ref{Fig.Na23.k1.cnls.MI.surf}--\ref{Fig.Na23.k1.n1.MI.surf}
show the maximum MI against the amplitudes (square root of the
particle density) of the $M_F=\pm 1$ and fields, for particular
differences in wave numbers ($k_1 - k_{-1} = 0, 1$ in dimensionless
units), for the different classes of cw solutions (CNLS, $n=0$, or
$n=1$), for $^{23}$Na (which is anti-ferromagnetic, with $c_2>0$) and
$^{87}$Rb (which is ferromagnetic, with $c_2<0$).  Quadratic Zeeman
splitting is zero in these figures.

\begin{figure}[hbt]
\centering\subfigure[]{\includegraphics[width=0.45\textwidth,angle=0]{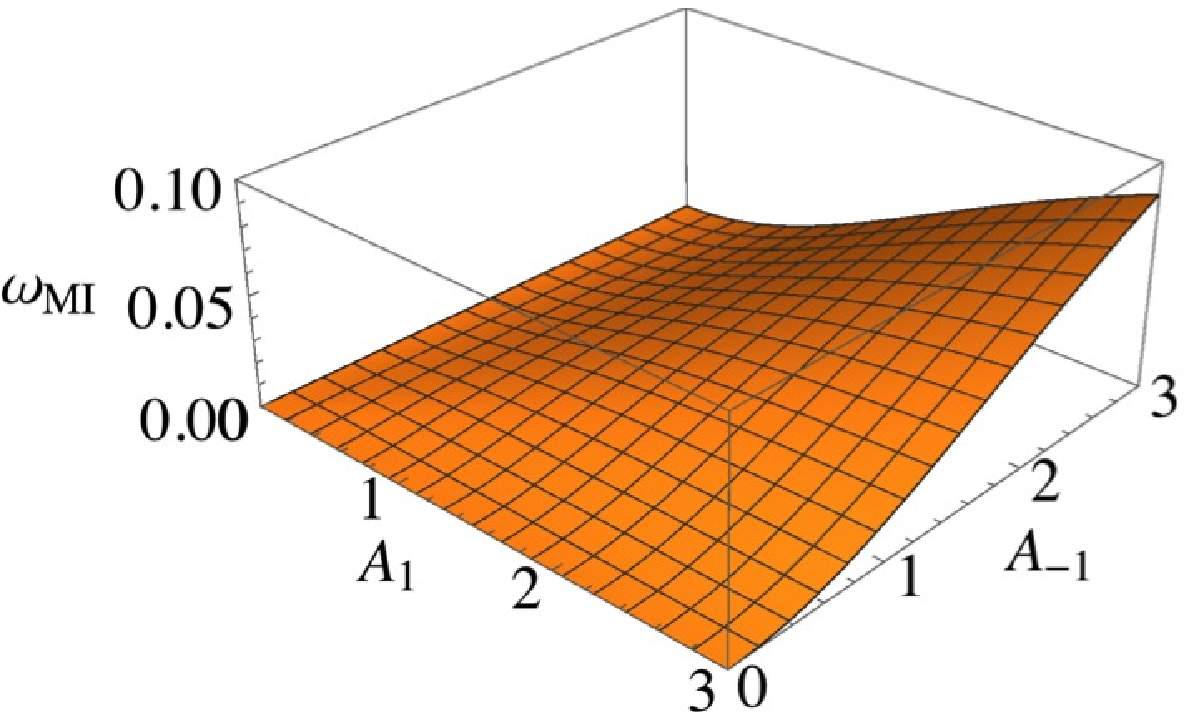}} 
\centering\subfigure[]{\includegraphics[width=0.45\textwidth,angle=0]{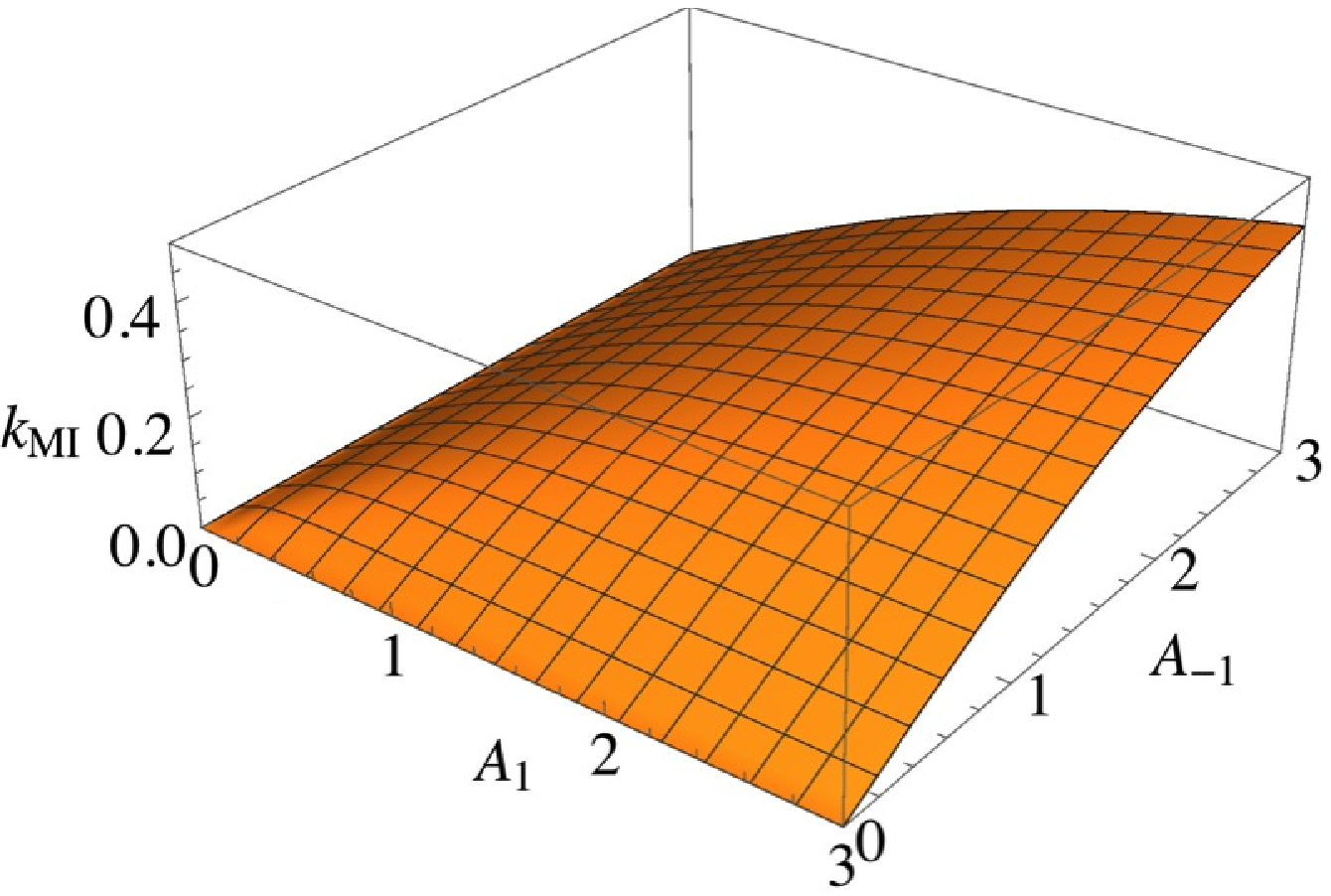}}
\caption{Maximum MI of the CNLS--type cws (zero $M_F=0$ particle
density) of a $^{87}$Rb BEC, as a function of the amplitudes of the
$M_F=\pm 1$ fields, where the spin components have identical wave
vectors $k_1=k_{-1}$, and quadratic Zeeman splitting is zero.  Part~(a)
shows the peak MI values, and the Part~(b) shows the wave vectors at
which the maxima occur.
All quantities in the figure are dimensionless; see Eqs.~(\ref{nondimensionalization}).}
\label{Fig.Rb87.k0.cnls.MI.surf}
\end{figure}

\begin{figure}[hbt]
\centering
\centering\subfigure[]{\includegraphics[width=0.45\textwidth,angle=0]{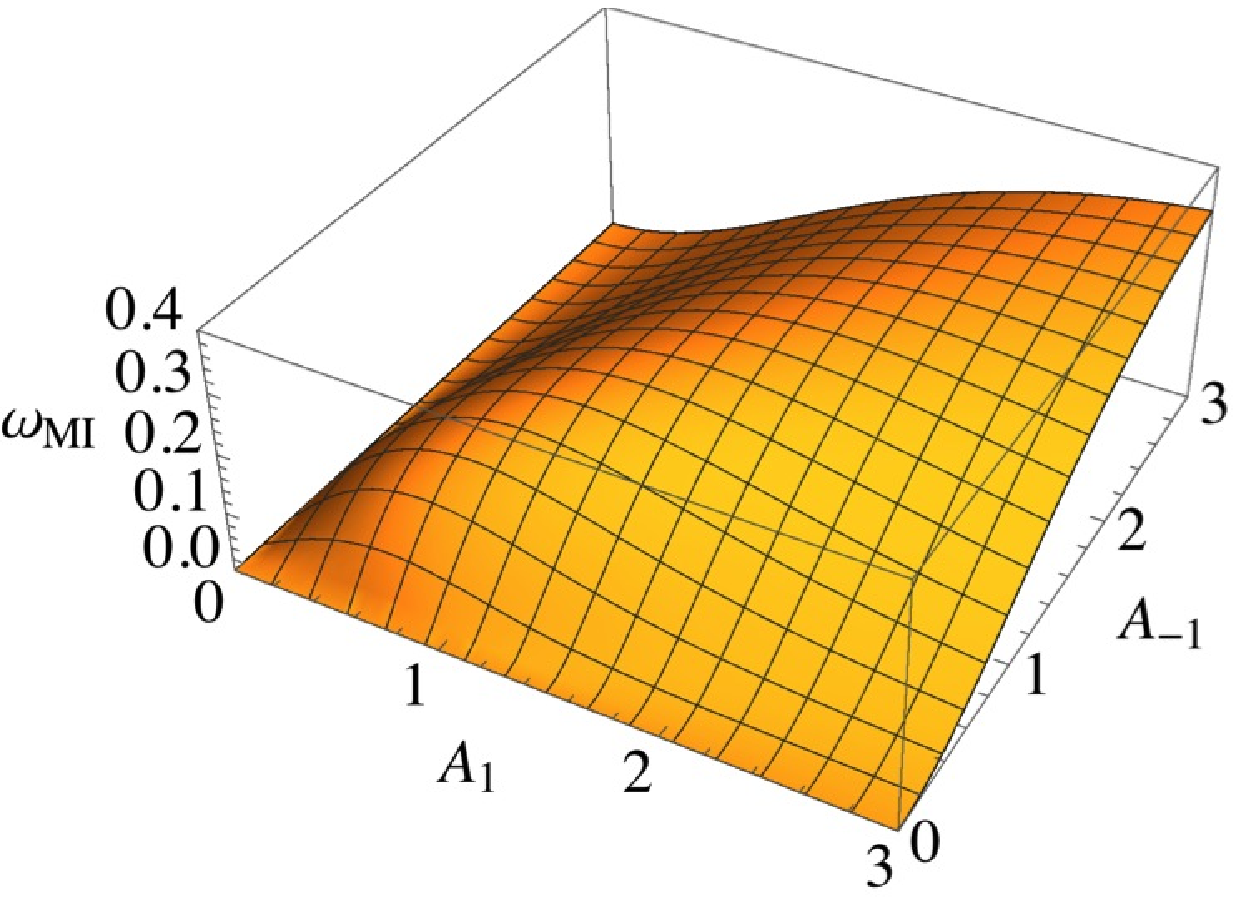}}  
\centering\subfigure[]{\includegraphics[width=0.45\textwidth,angle=0]{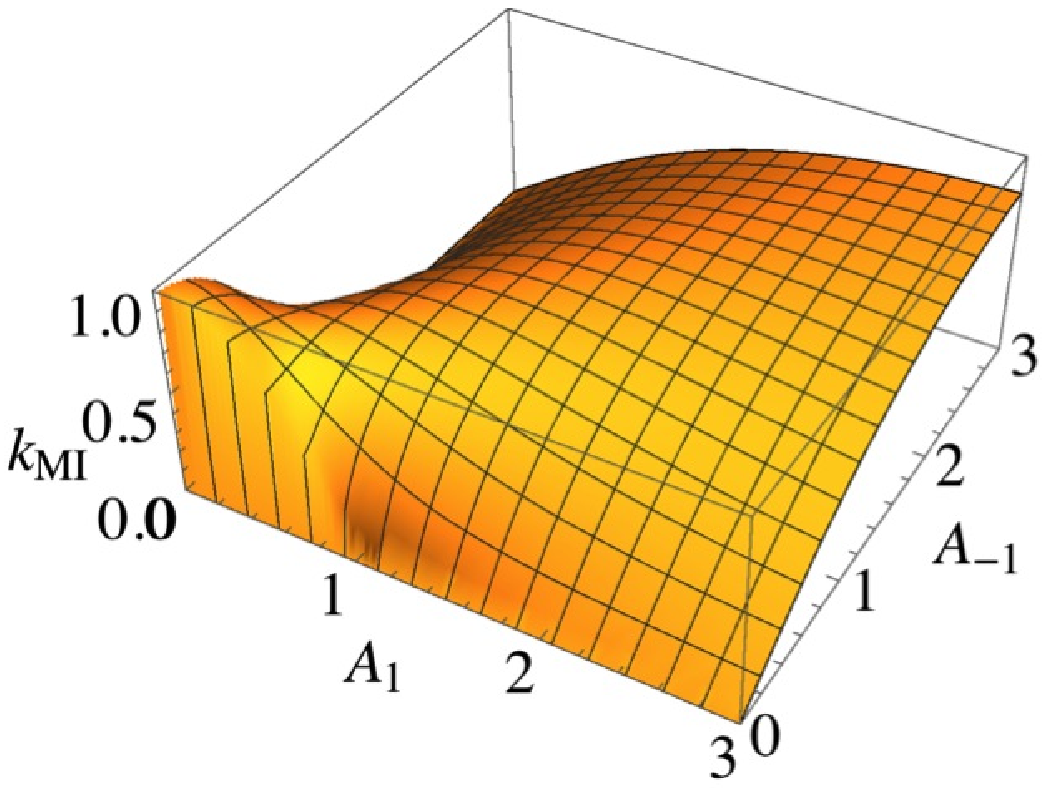}}
\caption{Maximum MI of the CNLS--type cws (zero $M_F=0$ particle
density) of a $^{87}$Rb BEC, as a function of the amplitudes of the
$M_F=\pm 1$ fields, where the spin components have
wave vectors that differ by $k_1-k_{-1}=1$, and quadratic
Zeeman splitting is zero.  Part~(a) shows the peak MI values, and
Part~(b) shows the wave vectors at which the maxima occur.
All quantities in the figure are dimensionless; see Eqs.~(\ref{nondimensionalization}).}
\label{Fig.Rb87.k1.cnls.MI.surf}
\end{figure}

Figure~\ref{Fig.Rb87.k0.cnls.MI.surf} shows the peak MI for the cws of
a $^{87}$Rb BEC with identical wave vectors in the spin $M_F=\pm 1$
components, $k_1 = k_{-1}$, and zero quadratic Zeeman splitting.  See
Fig.~\ref{Fig.sound.Rb87.k0.cnls} for the dispersion curves underlying
one point in this plot.
Figure~\ref{Fig.Rb87.k1.cnls.MI.surf} shows the peak modulational
instabilities for the cws of a $^{87}$Rb BEC with wave vectors in the
spin $M_F=\pm 1$ components with (in dimensionless units) unit
difference $k_1 - k_{-1} = 1$, and zero quadratic Zeeman splitting.
See Fig.~\ref{Fig.sound.Rb87.k1.cnls} for the dispersion curves
underlying one point in this plot.

Coupled NLS--type cws for a $^{23}$Na BEC with identical wave numbers
$k_1=k_{-1}$ and without quadratic Zeeman splitting have zero MI
(cf.\ Fig.~\ref{Fig.sound.Na23.k0.cnls}) for all values of the amplitudes
$A_1$, $A_{-1}$.
Figure~\ref{Fig.Na23.k1.cnls.MI.surf} shows the peak modulational
instabilities for the cws of a $^{23}$Na BEC with spin components with
a unit difference between the wave numbers of the spin $M_F=\pm 1$
components, $k_1 - k_{-1}=1$, and zero quadratic Zeeman splitting.
See Fig.~\ref{Fig.sound.Na23.k1.cnls} for the dispersion curves
underlying one point in this plot.

Next, let us show peak MI data for cross-sections of the parameter
space for the $n=0$ family of cw solutions, i.e., cws in which $A_0$,
the square root of the $M_F=0$ particle density, is as in
Eq.~(\ref{cw.A_0.solution}), with even-valued $n$.
The ($n=0$)--type cws for a $^{87}$Rb BEC with identical wave vectors in
the spin $M_F=\pm 1$ components, $k_1 = k_{-1}$, and zero quadratic
Zeeman splitting show vanishing MI at all values of the spin $M_F=\pm
1$ amplitudes $A_{\pm 1}$.  See Fig.~\ref{Fig.sound.Rb87.k0.n0} for
the dispersion curves underlying one point in the parameter space, and
note that the phonon band diagram is real-valued everywhere.
In contrast, the corresponding cw but with wave numbers that are not
all the same is subject to MI. Figure~\ref{Fig.Rb87.k1.n0.MI.surf}
shows the peak MI for cws of a $^{87}$Rb BEC with wave vectors in the
spin $M_F=\pm 1$ components with (in dimensionless units) unit
difference $k_1 - k_{-1} = 1$, and zero quadratic Zeeman splitting.
See Fig.~\ref{Fig.sound.Rb87.k1.n0} for the dispersion curves
underlying one point in this plot.

Next, consider the ($n=0$)--type cw for a $^{23}$Na BEC with identical
wave vectors in the spin $M_F=\pm 1$ components, $k_1 = k_{-1}$, and
zero quadratic Zeeman splitting.  The numerical spectral analysis
shows that the MI for all these cws is nil for all values of the spin
$M_F=\pm 1$ amplitudes $A_{\pm 1}$.  See
Fig.~\ref{Fig.sound.Na23.k0.n0} for the dispersion curves underlying
one point in this plot.

Figure~\ref{Fig.Na23.k1.n0.MI.surf} shows the peak MI for cws
of a $^{23}$Na BEC with a unit difference between the wave numbers of the spin
$M_F=\pm 1$ components, $k_1 - k_{-1}=1$, and zero quadratic Zeeman splitting.
See Fig.~\ref{Fig.sound.Na23.k1.n0} for the dispersion curves underlying one point in this plot.

Last in this series, let us show peak MI data for cross-sections of the parameter space for
the $n=1$ family of cw solutions, i.e., cws in which $A_0$, the square root
of the $M_F=0$ particle density, is as in Eq.~(\ref{cw.A_0.solution}), with odd $n$.
Figure~\ref{Fig.Na23.k1.n1.MI.surf} shows the peak MI for ($n=1$)--type cws
of a $^{23}$Na BEC with spin components
with a unit difference between the wave numbers of the spin $M_F=\pm 1$ components, $k_1 - k_{-1}=1$, 
and zero quadratic Zeeman splitting.
See Fig.~\ref{Fig.sound.Na23.k1.n1} for the dispersion curves underlying one point in this plot.

The figures are consistent with the well-known fact that there are no modulational
instabilities when there is only one spin component in an attractive BEC
\cite{Bogolubov.1947, BespalovTalanov.1966, Agrawal.2001}.

Figure~\ref{Fig.Rb87.k0.cnls.MI.surf} confirms that cws in $^{87}$Rb
without $M_F=0$ particles are modulationally unstable even for the
case in which all the wave numbers of the cws' spin components are the same.
Comparison with Fig.~\ref{Fig.Rb87.k1.cnls.MI.surf} shows that a difference in the
wave numbers of the $M_F=\pm 1$ fields increases the MI, mostly at low particle densities.
Differences in the wave numbers also increase the values of the wave numbers at which the MI is fastest.
At higher particle densities, the effects of the wave number differences become less important, and
the MI approaches the values of those of the cws with identical wave numbers in all spin components.

CWs in $^{23}$Na without $M_F=0$ particles are stable when all the wave numbers of the
spin components are the same (no surface plot of the maximum MI is displayed
because this is identically zero).
Figure~\ref{Fig.Na23.k1.cnls.MI.surf} shows that cw solutions of
$^{23}$Na without $M_F=0$ particles and with non-zero difference
in the wave vectors of the different spin components,
are modulationally unstable at low particle densities, but become stable at larger particle densities
For cw solutions with $M_F=0$ fields with relative phase corresponding to $n=0$,
MI is zero when the wave numbers of the spin components are all the same, 
MI is non-zero when the difference in the wave numbers is non-zero.
The effect of differences in the wave number is larger at low particle densities,
and smaller at high particle densities.
MI for a $n=1$ cw approaches zero when the densities of the spin $M_F=\pm 1$
components are almost the same (when the density of the $M_F=0$ particles in the cw
is greatest), and also when the densities of the spin $M_F=\pm 1$ components are
as different as is allowed (when the density of the $M_F=0$ particles in the cw is lowest).
It may be relevant for for experiments that small amounts of
particles with spin $M_F=0$ particles may destabilize a cw.

\section{Instability growth beyond the linear approximation}
\label{Sec:MI_evolution}

We carried out direct numerical simulations, forward in time, for cws
with initially small amounts of white noise. The full time evolution shows
a great many different phenomena, such as collision of phonons on top
of the cws and other effects when the system can no longer be considered
a perturbed cw. Phonon collisions will be analyzed in a separate article,
since to do so properly would require a great deal of space. The dynamics
in a system where a cw cannot be discerned is too large and varied a topic to
be contained within this article.
To better focus on confirmation of the spectral analyses, we show numerical
simulations with a spin-dependent nonlinear coefficient $c_2$ that is larger than the
physical values  in $^{23}$Na and $^{87}$Rb.  The fact that in these materials,
$c_2$ is smaller than $c_0$ by two orders of magnitude causes the MI to be weak.
With slow growth rates, once the unstable phonons grow to a certain amplitude,
they tend to collide with each other before growing very large,
and this obscures the MI-induced amplification.
Running simulations with larger values of $c_2$ allows us to avoid phonon collisions,
and to better focus on one phenomenon at a time: confirmation of the  MI.

\begin{figure}[hbt]
\centering
\includegraphics[width=8.5cm]{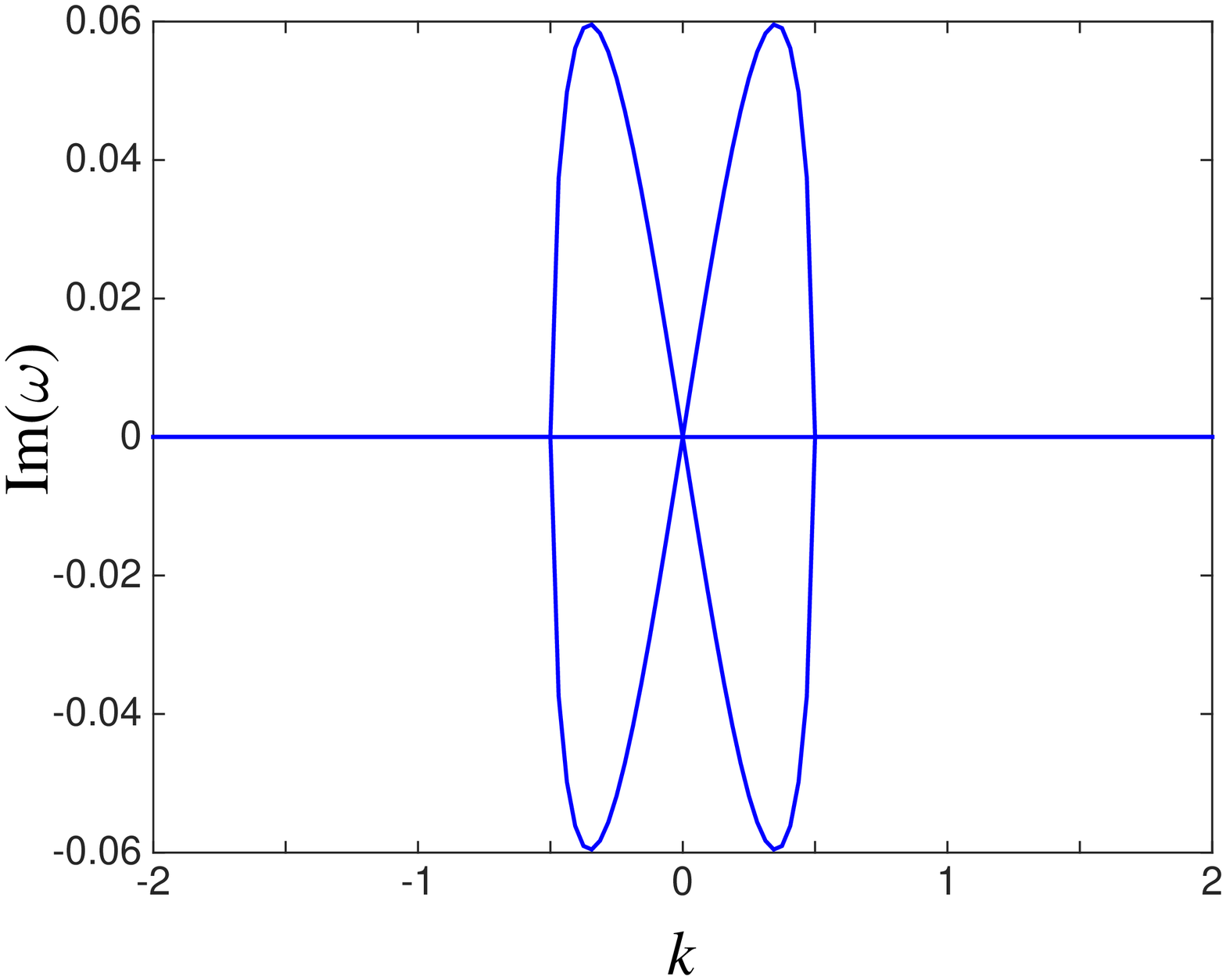} \\
\includegraphics[width=8.5cm]{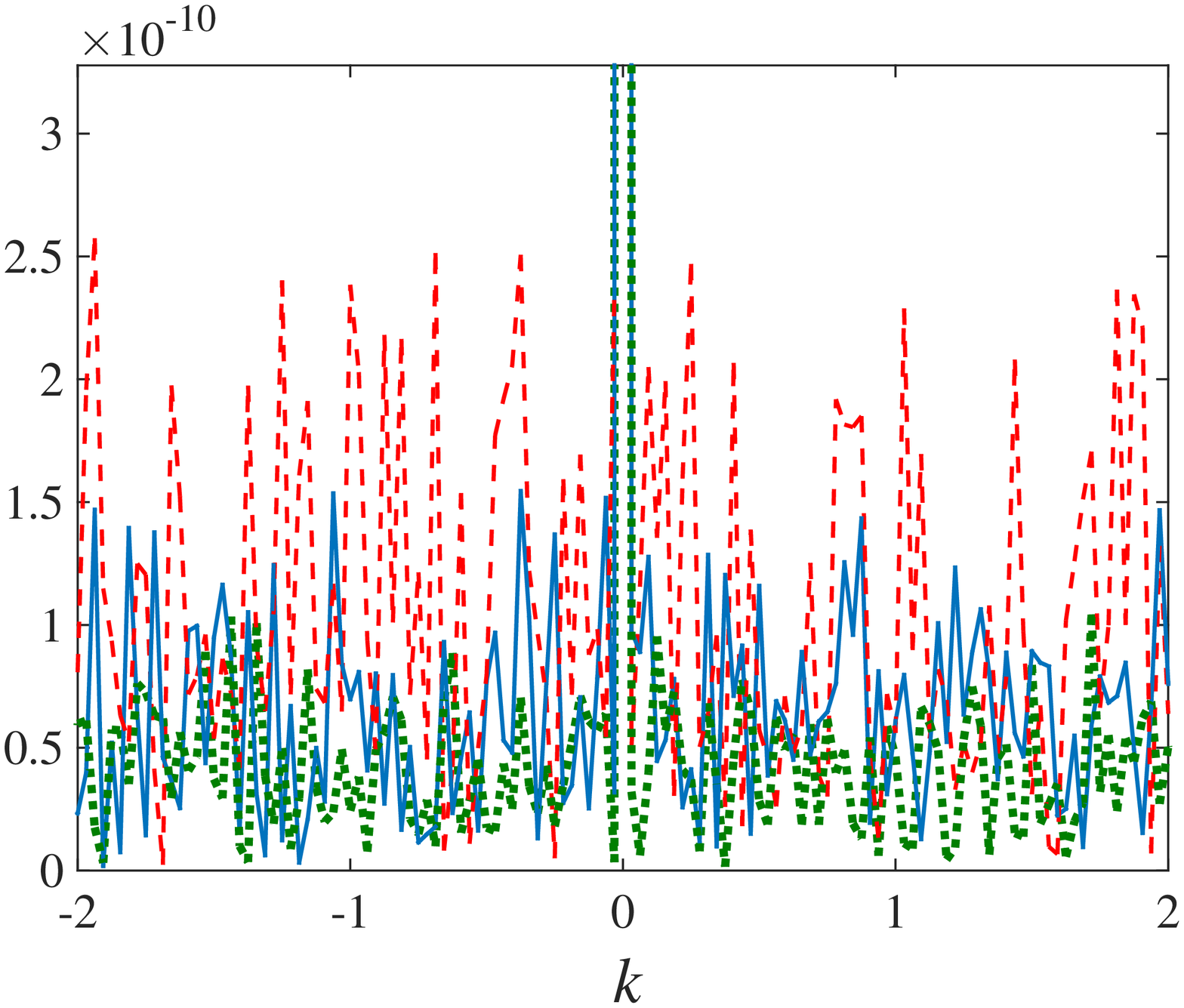}
\caption{(Color online) Part~(a) is the modulational instability band diagram for phonons
in a BEC with (dimensionless) parameters $\hbar=1$, $m=1$, $c_0=1$, $c_2=-.4793$,
on top of an ($n=0$)--type cw with
amplitudes $A_1=2.5$, $A_0=3.1826$, $A_{-1}=2.0$,
wave numbers $k_1=0.5$, $k_0=0$, $k_{-1}=-0.5$, 
and frequencies $\omega_1=10.6861$, $\omega_0=10.6722$, $\omega_{-1}=10.6583$.
Below is the spectral density at the start. The peak in the middle is the cw, and the
remainder of the spectrum is statistically flat.
All quantities in the figure are dimensionless; see Eqs.~(\ref{nondimensionalization}).}
\label{Fig.spinor_bec.band_diagram.MI}
\end{figure}

\begin{figure}[hbt]
\centering
\includegraphics[width=10cm]{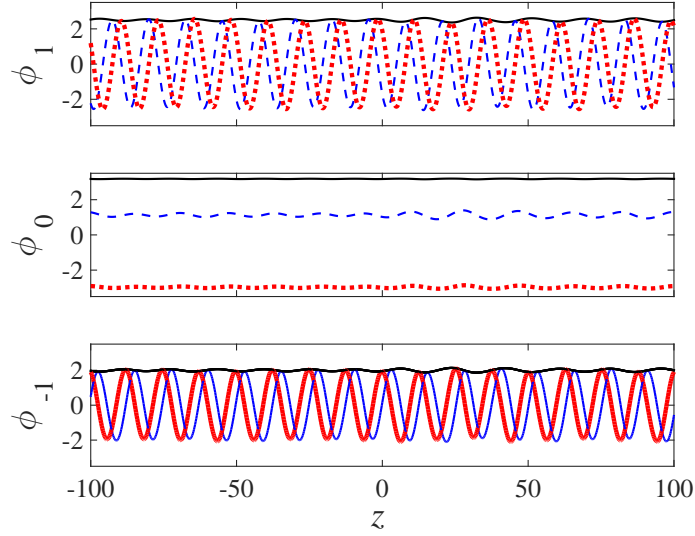}
\caption{(Color online) Amplitudes of the BEC spin fields $M_F=1,0,-1$, in dimensionless variables, at time $t=176.7.$ 
The magnitudes of the amplitudes are solid lines, the real parts dashed lines, and the imaginary parts dotted.
The cw structure is visible, and the amplified noise is also large enough to be visible to the eye.
All quantities in the figure are dimensionless; see Eqs.~(\ref{nondimensionalization}).}
\label{Fig.spinor_bec.phi.ReIm.z.t_177}
\end{figure}

\begin{figure}[hbt]
\centering
\centering\subfigure[]{\includegraphics[width=0.45\textwidth,angle=0]{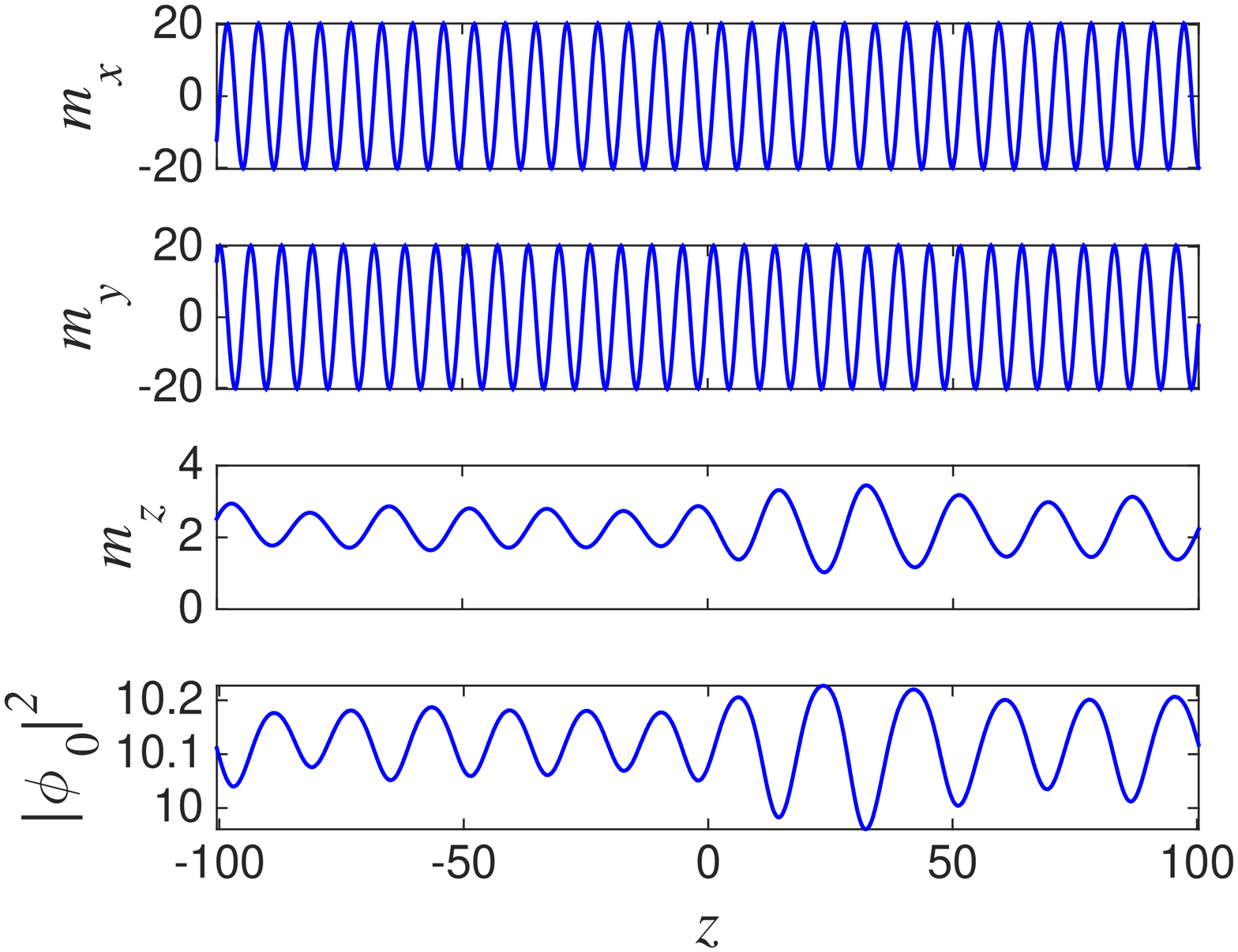}} 
\centering\subfigure[]{\includegraphics[width=0.45\textwidth,angle=0]{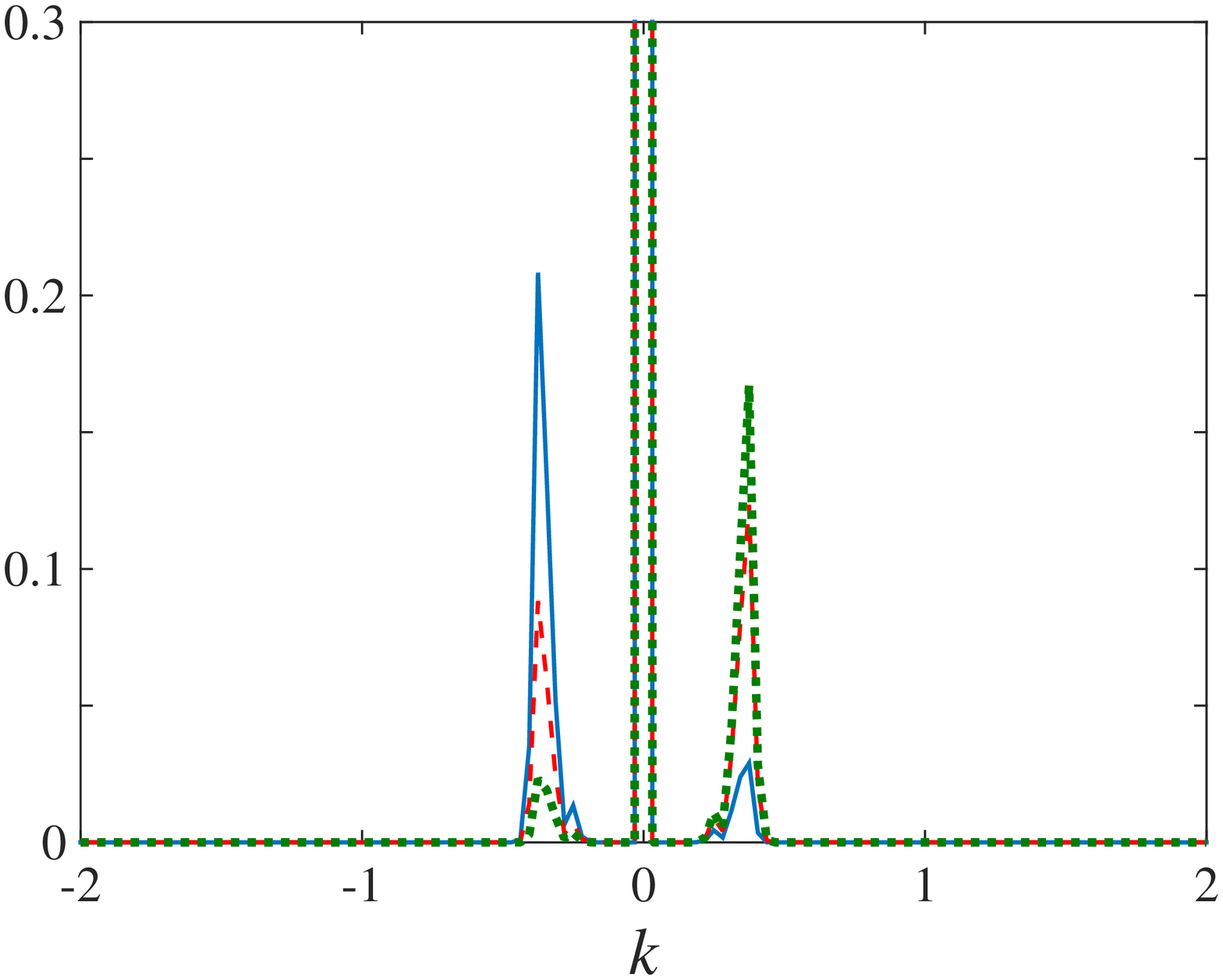}} 
\caption{(Color online) Snapshot at dimensionless time $t=176.7$.
Part~(a) shows the (dimensionless) magnetization components $(m_x, m_y, m_z)$ and
the particle density of the $m=0$ spin components, $|\phi_0|^2$. The pattern is the spin texture.
Part~(b) shows the squared amplitudes of the fields in momentum-space, i.e., the particle density as a function of wave number,
divided by the domain length.
The BEC field with spins $M_F=1,0,-1$ are represented by solid, dashed, and dotted lined, respectively.
All quantities in the figure are dimensionless; see Eqs.~(\ref{nondimensionalization}).}
\label{Fig.spinor_bec.processed_data.t_177}
\end{figure}

\begin{figure}[hbt]
\centering
\includegraphics[width=10cm]{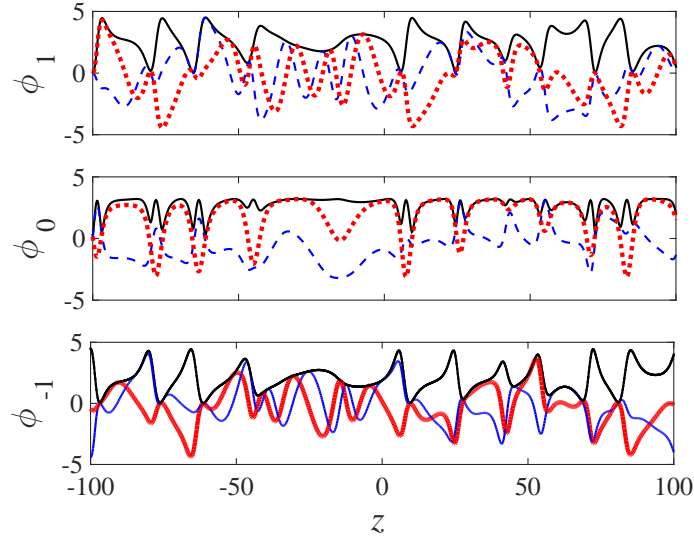}
\caption{(Color online)  Amplitudes of the BEC spin fields $M_F=1,0,-1$ at (dimensionless) time $t=372.9$. 
The magnitudes of the amplitudes are solid lines, the real parts dashed lines, and the imaginary parts dotted.
This is a snapshot of dynamical turbulence. The cw has been destroyed by amplified noise, and phonons are no longer helpful in
describing the system.
All quantities in the figure are dimensionless; see Eqs.~(\ref{nondimensionalization}).}
\label{Fig.spinor_bec.phi.ReIm.z.t_373}
\end{figure}

\begin{figure}[hbt]
\centering
\centering\subfigure[]{\includegraphics[width=0.45\textwidth,angle=0]{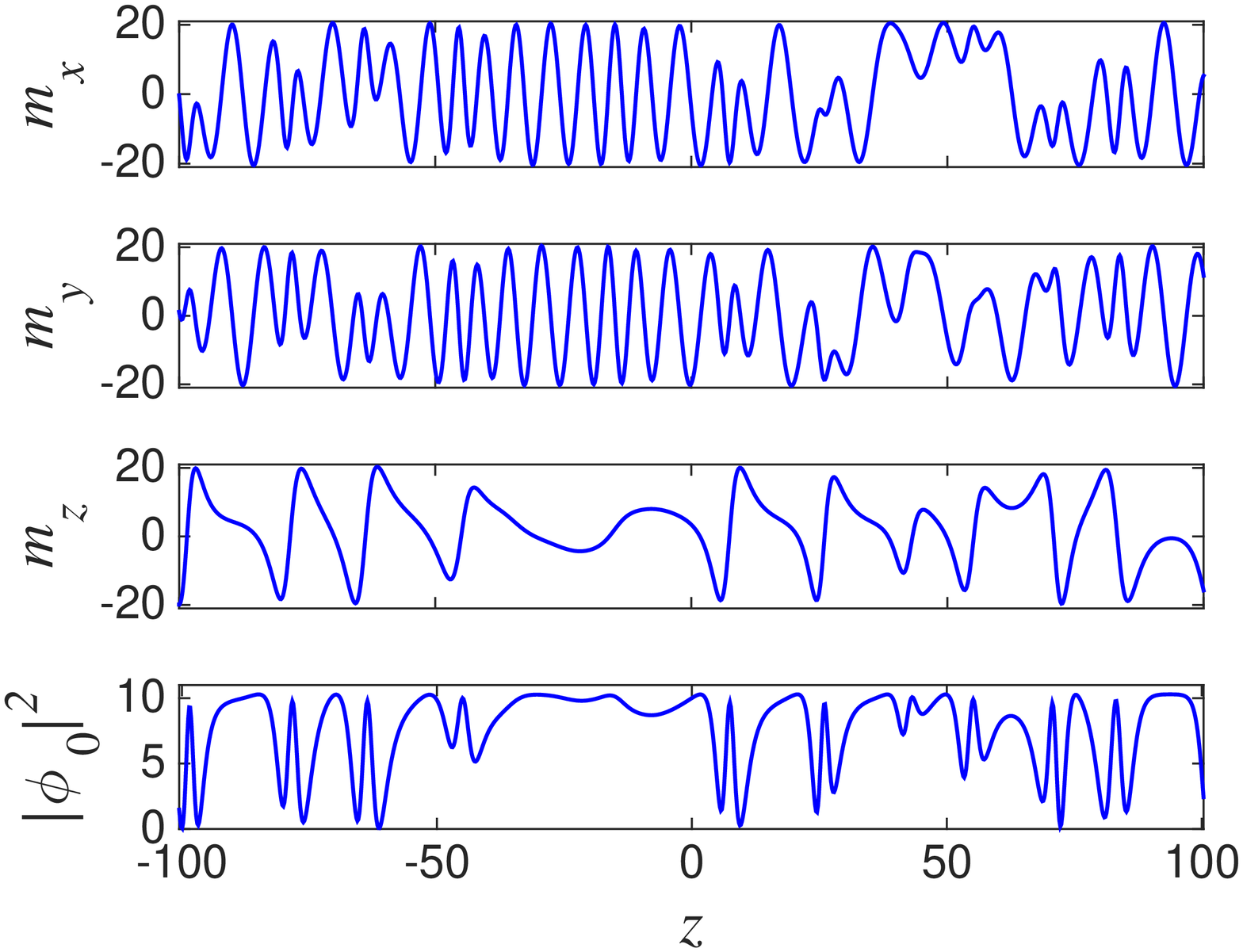}}
\centering\subfigure[]{\includegraphics[width=0.45\textwidth,angle=0]{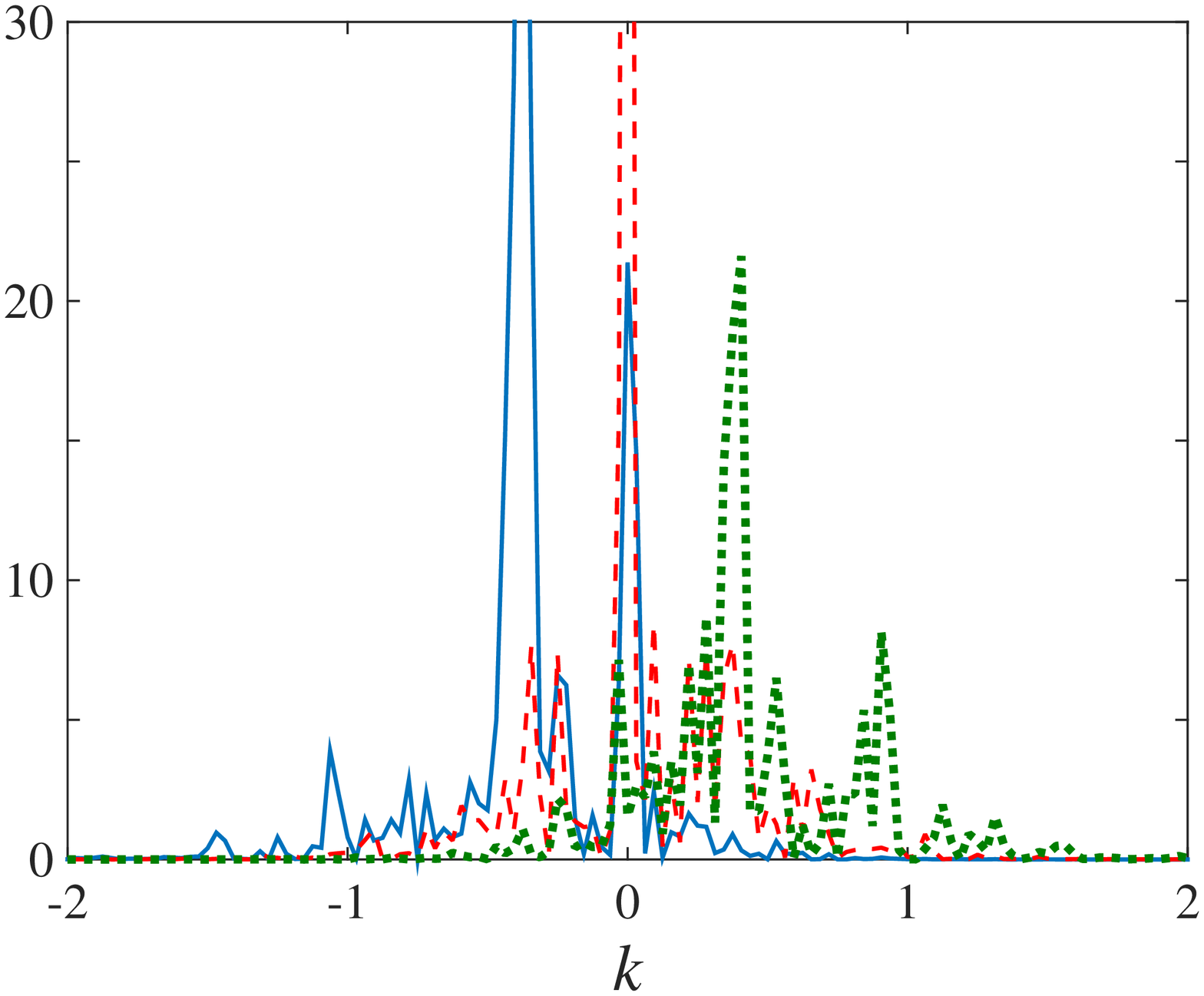}}
\caption{(Color online) Snapshot at dimensionless time $t=372.9$.
Part~(a) shows the (dimensionless) magnetization components $(m_x, m_y, m_z)$ and
the particle density of the $m=0$ spin components, $|\phi_0|^2$. The pattern is the spin texture.
Part~(b) shows the squared amplitudes of the fields in momentum-space, i.e., the particle density 
as a function of wave number, divided by the domain length. The BEC field with spins $M_F=1,0,-1$ 
are represented by solid, dashed, and dotted lines, respectively.
All quantities in the figure are dimensionless; see Eqs.~(\ref{nondimensionalization}).}
\label{Fig.spinor_bec.processed_data.t_373}
\end{figure}

Figures~\ref{Fig.spinor_bec.band_diagram.MI}-\ref{Fig.spinor_bec.processed_data.t_373}
show snapshots of a BEC with $c_2 = 100 c_2(^{87}\mathrm{Rb}) = -.4793 c_0$
that is initially an ($n=0$)--type cw with dimensionless amplitudes $A_1=2.5$, $A_0=3.1826$, $A_{-1}=2.0$,
wave numbers $k_1=0.5$, $k_0=0$, $k_{-1}=-0.5$, 
and frequencies $\omega_1=10.6861$, $\omega_0=10.6722$, $\omega_{-1}=10.6583$,
and initial weak white noise.
Figure~\ref{Fig.spinor_bec.band_diagram.MI} shows
the initial spectral particle densities of the fields with spin $M_F=1,0,-1$
and MI as a function of wave number.
At dimensionless time $t=176.7$, the noise at the modulationally
unstable wavelengths has grown, but not to the point that the BEC cannot still be considered as a perturbed cw.
Figure~\ref{Fig.spinor_bec.phi.ReIm.z.t_177} shows the amplitudes in real space---the magnitudes and the real and imaginary parts of the
fields $\phi_{1,0,-1}$. Figure~\ref{Fig.spinor_bec.processed_data.t_177}(a) shows the magnetization vector components and $|\phi_0|^2$, the density of particles of spin $m=0$.
The variation in space of the magnetization may be referred to as spin texture (cf.~\cite{Sadler.2006}).
Figure~\ref{Fig.spinor_bec.processed_data.t_177}(b) shows the spectral particle densities of the fields with spin $M_F=1,0,-1$.
At dimensionless time $t=372.9$, the noise has been amplified so much that the cw has been destroyed,
as one can see in Fig.~\ref{Fig.spinor_bec.phi.ReIm.z.t_373}, which shows the amplitudes (magnitudes and real and imaginary parts) in real space.
Figure~\ref{Fig.spinor_bec.processed_data.t_373}(a) shows the magnetization density vector (spin texture \cite{Sadler.2006}) 
and $|\phi_0|^2$, and Fig.~\ref{Fig.spinor_bec.processed_data.t_373}(b) shows the spectral particle densities of the spin components $M_F=1,0,-1$;
compare this with the initial noise and the MI spectrum in Fig.~\ref{Fig.spinor_bec.band_diagram.MI}.
The $t=176.7.$ snapshot looks similar to most of the prior development---a cw with amplified noise at the MI wavelengths---except for the
magnitude of the amplified noise.
At $t=372.9$, the MI has amplified some of the noise to the point that it has destroyed the cw.  The BEC is turbulent.
It is no longer meaningful here to talk about phonons.
The $t=372.9$ snapshot cannot be said to typify the fields past the point at which the cws have been destroyed;
there does not seem to be one typical static or statistical state of the BEC after destruction of the cw.
Comparison of Figs.~\ref{Fig.spinor_bec.processed_data.t_177} and~\ref{Fig.spinor_bec.processed_data.t_373}
shows that nontrivial textures (rather than a simple sinusoidal pattern) in the transverse components of the magnetization vector is
an indication that the underlying perturbed cw has been destroyed.
At lower values of $c_2$, phonon collisions can play a dominant role while the cw is still intact, 
and this can change the development of the noise. Effects of phonon collisions will be examined in another article.


\section{Summary and Conclusions}
\label{Sec:conclusions}

We examined the dynamics of sound waves (phonons, acoustic waves,
Bogoliubov excitations) in $F=1$ spinor BECs propagating on top of the
most general cw solutions. We focused more on cws with non-vanishing
$M_F=0$ spin components, since this does not have an analog in optics
and consequently has not been thoroughly investigated. Emphasis was
placed on $^{23}$Na (which is anti-ferromagnetic) and $^{87}$Rb
(which is ferromagnetic), both of which have repulsive nonlinearities.

At any given wave number, the phonons on top of a cw background can
take up to six distinct eigenvalues (frequencies or chemical
potentials), each with its own eigenvector (i.e., a specific mixture of
$M_F=1,0,-1$ spin components).  We showed the band diagrams (plots of
phonon frequencies against wave number) for representative cases of
each of the different supported types of cws, for $^{23}$Na and
$^{87}$Rb.  Many of the cws are modulationally unstable, i.e., have
frequencies with imaginary parts over some range of wave numbers.
Perhaps unexpectedly, the cws with non-vanishing $M_F=0$
components tend to be less subject to MI than cws with nil particle 
density for $M_F=0$, even though the Hamiltonian densities are higher
for the latter.  The MIs are in many cases weak and only occur for wave numbers
with magnitude up to a given point, beyond which there are no more instabilities
(or, equivalently, all phonons with wavelengths smaller than a certain value are stable).
Thus, even an ``unstable'' cw (unstable on an infinite domain), when
confined in a toroidal potential, may not support any unstable phonon modes.

Broadly speaking, differences in the wave numbers of the spin components 
$M_F=-1,0,1$ tend to make the cw unstable. All cws without any $M_F=0$ particles 
with nonzero wave number difference are subject to modulational instabilities,
even though such a cw in a $^{23}$Na BEC is stable when there is zero 
difference in the wave numbers. The destabilizing effects of a difference in 
wave numbers are significant when the particle densities are small,
and insignificant when the particle densities are large. 
All cws with $M_F=0$ particles with nonzero
wave number difference are subject to modulational instabilities,
even though the ($n=0$)--type cws in for both $^{23}$Na and $^{87}$Rb are stable 
when there is zero difference in the wave numbers.
Note that linear Zeeman splitting may be relevant here, even though it is 
mathematically trivial.  The transformation of variables that eliminates 
the linear Zeeman terms changes the frequencies and wave numbers of the 
components with $M_F=\pm 1$. Thus a cw in a BEC that feels a linear 
Zeeman splitting, if it is to have the same frequency (chemical potential) in all
original (not transformed) physical components, needs to have different 
wave numbers for the cw to exist. 

Our simulation of the dynamics of BECs confirmed the spectral
analyses, i.e., the band diagrams.  We observed that MI can lead to
exponential growth of noise, and that this can eventually destroy the
initial underlying cw and create a spin texture. Nonlinear
evolution of the phonons (which includes phonon collisions) is a very
large topic, which we will examine in greater detail in subsequent
work.

\begin{acknowledgments}
This work was supported in part by grants from the Israel Science
Foundation (No.~2011/295). 
\end{acknowledgments}

\end{document}